\documentclass[twocolumn, twocolappendix]{aastex63}
\pdfoutput=1 
\usepackage[T1]{fontenc}
\usepackage[utf8]{inputenc}
\usepackage[english]{babel}

\newcommand{\vtan}{$v_{\rm tan}$}

\newcommand{\cca}{Center for Computational Astrophysics, Flatiron Institute, Simons Foundation, 162 Fifth Avenue, New York, NY 10010, USA}

\usepackage[bottom,flushmargin]{footmisc}
\usepackage{amsmath}
\usepackage{todonotes}

\newcommand{\rcom}{$\rm R_{COM}$}
\newcommand{\vcom}{$\rm V_{COM}$}

\accepted{to ApJ \today}

\shorttitle{M31's Center of Mass Evolution Owing to M33}
\shortauthors{Patel, Garavito-Camargo \& Escala}
\graphicspath{{./}{figures/}}

\defcitealias{patel17a}{P17}
\defcitealias{patel23}{P23}

\begin{document}

\title{The M31-M33 Interaction: Impact on M31's Center of Mass Motion and Satellite Orbits}

\correspondingauthor{Ekta Patel}
\email{ekta.patel@utah.edu}

\author[0000-0002-9820-1219]{Ekta~Patel}\thanks{NASA Hubble Fellow}
\affiliation{Department of Physics and Astronomy, University of Utah, 115 South 1400 East, Salt Lake City, Utah 84112, USA} 

\author[0000-0002-0786-7307]{Nicol\'as Garavito-Camargo}
\affiliation{\cca}

\author[0000-0002-9933-9551]{Ivanna Escala}
\affiliation{Space Telescope Science Institute, 3700 San Martin Drive, Baltimore, MD 21218, USA}

 
\begin{abstract}
Inspired by recent studies of the Milky Way--LMC interaction and its implications for the Milky Way's global dynamical history, we investigate how the massive satellite galaxy M33 influences Andromeda's (M31) center of mass (COM) position and velocity as it passes through M31's halo. Using recent 6-dimensional phase space measurements for both galaxies, we use backward integration to revisit M33's orbital history in a massive M31 potential ($3\times10^{12}\,M_{\odot}$) for the first time. As previously concluded, we find that a first infall orbit is still the most statistically significant ($\gtrsim$ 90\%) orbital solution for M33, except for a high mass M31 combined with M31 proper motions from HST (as opposed to \textit{Gaia}), where there is a greater likelihood ($\sim$65 \%) of a previous encounter. However, the minimum distance between M33 and M31 during this passage is typically $\geq$ 100~kpc, two to three times larger than the distance required to explain M33's warped stellar and gaseous disks. We quantify the magnitude and direction of M31's evolving COM position (\rcom{}) and velocity (\vcom{}) owing to M33, finding \rcom{}$\approx$100-150~kpc at maximum and \vcom{}$\approx$20-40 km s$^{-1}$. Furthermore, we explore the implications of this phenomenon for the M31 satellite system, specifically whether M33's gravitational influence is linked to the lopsided distribution of M31 satellites and whether M33 significantly perturbs the orbits of other M31 satellites. While M33 alone may not explain the lopsided nature of M31's satellite system, its dynamical impact is non-negligible and must be accounted for in future dynamical studies of the M31 system.
\end{abstract}


\section{Introduction}\label{sec:intro}

Owing to six-dimensional (6D) phase space information measurements (3D position and 3D velocity) for galaxies in the Local Group (LG), we can now rigorously study the past orbital trajectories of nearby galaxies. This data has been especially useful in understanding host-satellite interactions for the satellites orbiting around the Milky Way (MW) and Andromeda (M31). With precise orbital information, it is possible to determine the interaction histories of satellite galaxies \citep[e.g.,][]{kallivayalil13, sohn17, battaglia22, Magnus22}, the external dynamical processes that may have shaped their morphologies \citep[e.g.,][]{patel17a, sohn20, bennet24}, and the properties of $\Lambda$CDM dark matter halos that satellite galaxies orbit within \citep[e.g.,][]{patel18a}. 

Recent studies of the Large Magellanic Cloud (LMC) and the MW have shown that the passage of the LMC through the MW's halo can induce a displacement and an apparent velocity offset from the vantage point of the MW's disk, or reflex motion, on the center of mass (COM) of the MW \citep{petersen21, erkal21,  Yaaqib24, chandra24, bystrom24}. These phenomena are largely due to the high total mass of the LMC \citep[$M_{\rm LMC} > 10^{11}\, M_{\odot}$; e.g.,][]{erkal20,vasiliev21, shipp21, watkins24}, which is nearly 10-20\% the total mass of the MW. 

\citet{gomez15} first used rigid halo models and backward orbit calculations to conclude that the LMC can displace the COM of the MW by as much as 30~kpc in the last 0.3-0.5 Gyr and additionally impart a velocity offset of 75 km s$^{-1}$. Subsequently, the orbital barycenter of the MW-LMC system also shifts away from the MW's present-day COM by approximately 14~kpc. More sophisticated models of the MW-LMC system using idealized N-body simulations have further refined the expected velocity offset and displacement of the MW with live halos \citep{garavitoc19, cunningham20, petersen20}. \citet{garavitoc19} showed the MW's disk and inner halo ($\rm r_{GC} <$ 30~kpc) are displaced by 20-50~kpc with respect to the outer halo, and this displacement changes as a function of Galactocentric radius. Similarly, the reflex motion also changes with radius and is largest at $\rm r_{GC} >$ 30~kpc. These claims have been confirmed via observations\footnote{A measurement of velocity offset between the inner and outer MW halo is often referred to as travel velocity.}. For a recent detailed discussion on measurements of the MW reflex motion, we refer readers to \citet{bystrom24}.

The resulting MW COM displacement and reflex motion induced by the LMC have been shown to have major consequences for the distribution and kinematic properties of other MW substructures, including stellar streams, globular clusters, and satellite galaxies \citep{patel20, vasiliev21, battaglia22, Magnus22, lilleengen23, Brooks24a}. For example, \citet{gc24} showed using idealized N-body simulations of the MW-LMC and analogs from the Latte suite of FIRE simulations that the passage of a massive satellite on a close pericentric passage can indeed cause alignment in the orbital poles of satellite galaxies owing to the two-body interaction of the host-satellite. 
Forthcoming work will explore the full response of the MW's halo to the LMC on the orbits of all known satellite galaxies around the MW  with measured 6D phase space information (Garavito-Camargo et al., in prep.; Patel et al., in prep.). 

COM displacements are common in and expected in cosmological simulations, as shown by \citet{salomon23}. This analysis quantified the COM of host galaxies and their satellites in the HESTIA simulations, a high-resolution suite designed to mimic the LG. From \citet{salomon23}, we expect velocity shifts of order 10 km s$^{-1}$ in the absence of massive satellites and $\sim$60 km s$^{-1}$ when there are massive satellites. In \citet{gc24}, we showed that in the Latte suite, massive satellites can induce velocity shifts up to 80 km s$^{-1}$ shortly after pericenter. Overall the shifts depend on the mass ratio of the host and satellite and on the pericenter distance, further motivating the need to study this effect in M31's halo.

In this work, we turn to the MW's nearest neighbor -- Andromeda (M31) -- the only massive galaxy that can currently be studied at a comparable level of detail to the MW. M31 also hosts a massive satellite galaxy, Triangulum (M33), that is just slightly more massive than the LMC \citep[M33 $M_{*} \approx 3.2 \times 10^9\,M_{\odot}$;][]{corbelli03, guo10}. Recent mass estimates for M31 yield upper limits of $2.5-3\times10^{12}\, M_{\odot}$ \citep{patel23}. Thus, M31-M33 and MW-LMC both have a mass ratio of $\approx$1:10. As such, it is expected that M33 may also induce a dynamical response from M31, which may, in turn, affect the phase space distribution of other M31 substructures. Moreover, M31's assembly history is proposed to be quite active compared to that of the MW \citep{mcconnachie18}. It may have recently undergone a major merger \citep{hammer18, dsouza18}, and also hosts a bright satellite, M32, in front of its disk whose interaction history with M31 is still unknown. Additionally, the properties of M31 satellites differ from the overall MW satellite population. For example, M31's satellites exhibit systematically different quenching times compared to MW satellites, and this may be a result of their different accretion histories \citep{weisz19}. Thus, a detailed analysis of another satellite system can illuminate whether the conclusions drawn from the dynamics of the MW-LMC interaction are also present in external satellite systems \citep[see also][]{salomon23}. 

Here, we aim to quantify the magnitude and direction of M31's evolving COM position and velocity as a result of M33's passage. This exercise is analogous to \citet{gomez15}, who used a similar orbital setup to determine the consequences of artificially fixing the MW's COM as the LMC approaches the MW on first infall. 

In this paper, we compute orbits using M31 proper motions measured with \emph{Gaia} eDR3 \citep{salomon21} and the \emph{Hubble Space Telescope} \citep[HST;][]{vdm12ii} as these are the two most precise measurements in the literature from two different observatories. We also consider the consequences of two different M31 masses, including recent high mass estimates for M31 \citep[$\sim3\times10^{12}\, M_{\odot}$;][]{patel23} and low mass M31 estimates from the literature \citep[e.g.,][]{diaz14, penarrubia14, patel17b, carlesi22}. Ultimately, we aim to determine how the dynamical impact of M33 might influence the global dynamics of the M31 system. Recent work presenting improved distances to satellites around M31 \citep{savino22} has shown that $\approx$80\% of known satellites are in an asymmetric distribution on the near side of M31. We also investigate whether the gravitational influence of M33 could help explain this spatial distribution.

This paper is organized as follows. Section \ref{sec:data} summarizes the two observational data sets used to generate orbits for M31-M33, the orbital models, and the integration scheme. Section \ref{sec:orbit_results} presents orbital histories and uncertainties for M33 in a low mass and high mass M31 potential. In Section \ref{sec:reflex}, we quantify M31's evolving COM position and velocity. Section \ref{sec:discussion} includes an investigation of whether M33 is linked to the observed distribution of M31 satellites and the prospects of observing the reflex motion of M31 using stellar tracers. Finally, we conclude and summarize in Section \ref{sec:conclusions}.

\begin{deluxetable*}{lcccccccc}[t]
\tablecaption{M31 and M33 Position and Velocity Vectors}
\label{tab:coords}
\tablewidth{0pt}
\tablehead{
\colhead{Galaxy}  & \colhead{PM ref.} & \colhead{$\rm D_{MW}$} &  \colhead{$x$} & \colhead{$y$} & \colhead{$z$} &  \colhead{$v_x$} & \colhead{$v_y$} & \colhead{$v_z$} \\
\multicolumn1c{} & 
\colhead{} & 
\colhead{(kpc)} & \colhead{(kpc)} &  \colhead{(kpc)} & \colhead{(kpc)} &  \multicolumn1c{(km $s^{-1}$)} & \multicolumn1c{(km $s^{-1}$)}  & \multicolumn1c{(km $s^{-1}$)} 
}
\startdata
M31 & HST+sats & $770.0^{+40}_{-40}$ & -378.9 $\pm$ 10.6 & 612.7 $\pm$ 17.5 & -283.1 $\pm$ 8.1 & 66.1 $\pm$ 26.7 & -76.3 $\pm$ 19.0 & 45.1 $\pm$ 26.5 \\
M31 & \emph{Gaia} eDR3 & $776.2^{+22}_{-21}$ & -382.5 $\pm$ 10.7 & 617.6 $\pm$ 17.6 & -284.4 $\pm$ 8.1 & 34.9 $\pm$ 33.5 & -129.3 $\pm$ 23.1 & -13.7 $\pm$ 27.7\\
M33 & VLBA & $859.0^{+24}_{-23}$ &  -515.4 $\pm$ 14.8 & 531.3 $\pm$ 15.5 & -445.3 $\pm$ 13.1 & 39.4 $\pm$ 23.0 & 94.0 $\pm$ 25.1 & 144.0 $\pm$ 30.7\\
\enddata
\tablecomments{Galactocentric three-dimensional position and velocity vectors for M31 and M33. Uncertainties are computed in a Monte Carlo fashion by drawing samples from the $1\sigma$ error space of distance, proper motion, and line-of-sight velocity. Distances are adopted from \citet{savino22} and \citet[][see also references therein]{vdmG08}. Line-of-sight velocities are from \citet{slipher13} and \citet{corbelli97} for M31 and M33, respectively. Proper motions are adopted from the following references:  \citet{vdm12ii} -- HST+sats (M31), \citet{salomon21} -- \emph{Gaia} eDR3 (M31), and \citet{brunthaler05} -- VLBA (M33).}
\end{deluxetable*}

\section{Data and Methods}
\label{sec:data}
Integrating the orbital history of the M33-M31 system requires 6D phase space information for both galaxies. These coordinates are derived from a combination of observational data, including distances, line-of-sight velocities, and proper motions from the recent literature. Here, we discuss the specific values adopted for M31 and M33.

\subsection{Observed Data}
\subsubsection{M31}
We adopt two sets of 6D phase space information for M31. The first is based on a distance of $\rm D=770\pm40\,~kpc$ \citep[see][and references therein]{vdmG08}, a line-of-sight velocity of $\rm v_{LOS}=-301 \, km \, s^{-1}$ \citep{slipher13}, and HST-based proper motions from \citet{vdm12ii}, ($\mu_{\alpha *},\mu_{\delta}$) =  (34.30$\pm$8.44, -20.22$\pm$7.78) $\mu$as yr$^{-1}$. These proper motions are the weighted average of direct measurements from HST and the indirect proper motion implied by the motions of M31 satellite galaxies. The 6D phase coordinates resulting from this combination of measured data are listed in the first row of Table \ref{tab:coords} and will be denoted as the HST+sats data set. These 6D phase space coordinates are identical to those adopted in \citep[][hereafter \citetalias{patel17a}]{patel17a} for M31.

The second set of M31 6D phase space data is derived from the combination of a distance $\rm D=776.2^{+22}_{-21}\,~kpc$ \citep{savino22}, $\rm v_{LOS}=-301 \, km \, s^{-1}$ \citep{slipher13}, and \emph{Gaia} eDR3-based proper motions from young blue main sequence stars, ($\mu_{\alpha *},\mu_{\delta}$) =  (48.98$\pm$10.47, -36.85$\pm$8.03) $\mu$as yr$^{-1}$,  \citep{salomon21}\footnote{\citet{salomon21} also report an M31 proper motion measured with red AGBs and giant stars, however, this sample is likely susceptible to MW foreground contamination and therefore yields a significantly different M31 proper motion than the fiducial blue sample.}. The 6D phase space coordinates derived from these measurements are provided in the second row of Table~\ref{tab:coords} and will be denoted as the \emph{Gaia} eDR3 data set. 

The main difference between these data sets for M31 is that the latter set of \emph{Gaia} eDR3-based proper motions yields a tangential velocity of $\rm v_{tan} = 79.8\pm 38.3 \, km \, s^{-1}$ compared to just $\rm v_{tan} = 17 \, km \, s^{-1}$ (with a $1 \sigma$ confidence region of $\rm v_{tan} \leq 34.3 \, km \, s^{-1}$) for the HST+sats data.

Note that the proper motions from \citet{salomon21} are consistent with the weighted average of HST and \emph{Gaia} DR2 results reported in \citet{vdm19}. The more recently measured M31 proper motions from \citet{rusterucci24}\footnote{We refer readers to Table 1 of \citet{rusterucci24} for a summary of M31 proper motion measurements.} are even more similar to those reported in \citet{salomon21}. The results of this work are not expected to differ significantly if we adopt either the \emph{Gaia} DR2 \citep{vdm19} or \emph{Gaia} DR3 \citep{rusterucci24} results.

\subsubsection{M33}
\citet{brunthaler05} first measured the proper motion of M33 by tracking water masers using the \textit{Very Long Baseline Array} (VLBA), reporting ($\mu_{\alpha *},\mu_{\delta}$) =  (23$\pm$7, 8$\pm$9) $\mu$as yr$^{-1}$. Other measurements for M33's proper motion are from \emph{Gaia} DR2 \citep{vdm19} and DR3 \citep{rusterucci24}. The \emph{Gaia} values are in agreement with one another and with the VLBA values at the 2$\sigma$ level, but the corresponding uncertainties are at least a factor of three or more larger with \emph{Gaia} than those of the VLBA measurements. Therefore, we adopt the VLBA measurement in this work, a distance of $\rm D=859^{+24}_{-23}\,~kpc$ \citep{savino22}, and a line-of-sight velocity of $\rm v_{LOS}=-179.2 \, km \, s^{-1}$ \citep{corbelli97}. The resulting 6D phase space coordinates for this combination of measured data are listed in the last row of Table~\ref{tab:coords}.

\subsection{Orbital Methods}
\label{subsec:orbit_methods}
\subsubsection{Galaxy Potentials}

We adopt two mass models for M31, including a high mass M31 ($\rm M_{vir}= 3 \times 10^{12} \, M_{\odot}$) and a low mass M31 ($\rm M_{vir}= 1.5 \times 10^{12} \, M_{\odot}$) to compare to previous work. These halo masses are selected to align with recent M31 mass estimates in the literature \citep[see compilation in][]{patel23}. 

Following the methods in \citetalias{patel17a} and \citet{sohn20}, we build M31 potentials using three components. The halo is an NFW profile \citep{nfw96}, the disk is a Miyamoto-Nagai potential \citep{mn75}, and the bulge is a Hernquist sphere \citep{hernquist90}. The halo is additionally adiabatically contracted using the \texttt{CONTRA} code \citep{contra}, and the dark matter halo profiles are truncated at the virial radius. Note that for the NFW halo, $\rm M_{halo} = M_{vir} - M_{disk} - M_{bulge}$.

M31 parameters for both the low and high mass potentials are listed in Table~\ref{table:params}. The low mass M31 potential has the same virial mass as the low mass M31 potential considered in \citetalias{patel17a}, but different bulge and disk parameters (see Figure~\ref{fig:p17a_comp}). We compare the resulting M33 orbit in both the old and new low mass M31 potential in Appendix~\ref{sec:appendixA}. 

In this work, the disk and bulge parameters are chosen to match those provided in \citet{tamm12, courteau11} and to simultaneously match the rotation curve from \citet{corbelli10}, which peaks at a circular velocity of $\rm \approx 250 \, km s^{-1}$. We note the resulting disk mass adopted in this work agrees with that estimated from observations, which is $9 \pm 2 \times 10^{10} \, M_{\odot}$ of surviving stellar material \citep{williams17}. Recent M31 disk height measurements report $z_d = 770 \pm 80$ pc, which is also consistent with our choice at the 2$\sigma$ level \citep{dalcanton23}. We exclude any contributions from M31's gaseous disk in this work as its total gas mass is estimated to be less than 10\% of its stellar mass \citep{chemin09, corbelli10}.

\begin{figure*}[t]
    \centering
\includegraphics[width=0.85\textwidth, trim=0mm 0mm 0mm 20mm]{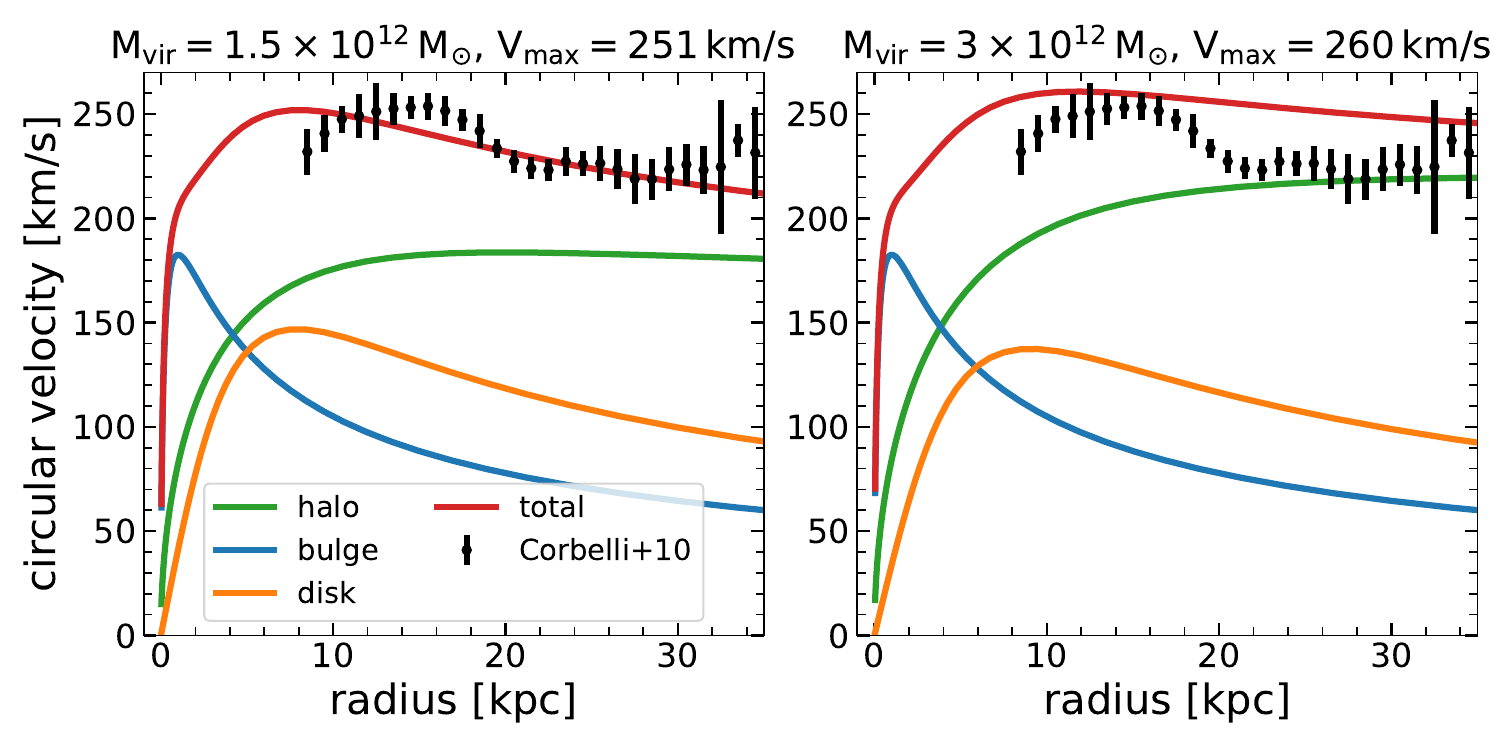}
    \caption{Rotation curves for the low (left) and high mass (right) M31 potentials models. In each panel, the individual circular velocity profiles are indicated for the halo (green), disk (orange), and bulge (blue) components. The total rotation curve is shown in red. The black data points are those reported in \citet{corbelli10} from HI observations. The parameters used to construct these models are listed in Table~\ref{table:params}. Both M31 models reach peak velocities of $\rm v_{circ} \sim$ 250-260 km s$^{-1}$, in agreement with the observed data.}
    \label{fig:rcs}
\end{figure*}

Rotation curves for both the low and high mass M31 potentials are illustrated in Figure~\ref{fig:rcs}, which shows the individual circular velocity profiles for the disk, bulge, and halo, as well as the total rotation curve compared to observational data from \citet{corbelli10}. The low mass M31 potential exhibits circular velocities very consistent with those from observations as indicated by the intersection of the black data points and the red line in the left panel of Figure~\ref{fig:rcs}. We note that the high mass M31 potential has systematically higher circular velocities than the values derived from the HI observations by \citet{corbelli10}.

\begin{table}[t]
\centering
\caption{Parameters used for the analytic representation of M31's potential in our models. Two mass models are adopted, corresponding to a low mass and a high mass M31.}
\label{table:params}
\begin{tabular}{lcc}\hline\hline 
 & Low Mass M31 & High Mass M31 \\ \hline
$\rm M_{vir}$ [10$^{10}$ $\rm M_{\odot}$] & 150 & 300 \\ 
c$_{\rm vir}$ & 9.56 & 9.08 \\ 
$\rm R_{vir}$ [kpc] & 299 & 377 \\
M$_{\rm d}$ [10$^{10}$ $\rm M_{\odot}$] & 7.3 & 7.3 \\ 
R$_{\rm d}$ [kpc] & 5.0 & 5.8 \\
z$_{\rm d}$ [kpc]  & 0.6 & 0.6 \\ 
M$_{\rm b}$ [10$^{10}$ $\rm M_{\odot}$] & 3.1 & 3.1  \\ 
R$_{\rm b}$ [kpc] & 1.0 & 1.0 \\ \hline
\end{tabular}
\end{table}

As in \citetalias{patel17a}, M33 is modeled with just one component -- a dark matter halo -- at three different virial infall masses. These masses are $\rm 5, 10, 25 \times 10^{10} \, M_{\odot}$ and were determined using abundance matching models \citep[e.g.,][]{guo11} assuming a stellar mass of $3.2 \times 10^9\,M_{\odot}$ \citep{corbelli03,guo10}. M33 is modeled as a Plummer sphere \citep{Plummer11}, and the corresponding scale lengths for these virial masses are $k_{M33}$ = 1, 11.5, and 21~kpc. These scale lengths are determined by matching the dynamical mass of $\rm M_{dyn}(<15\,~kpc) \gtrsim 5 \times 10^{10} \, M_{\odot}$ as determined by \citet{corbellisalucci}.

Dynamical friction is modeled following the methods of \citetalias{patel17a}, and identical parameters are used for the Coulomb logarithm. In this setup, M33 experiences dynamical friction as it passes through M31's halo. M31 also experiences the gravitational force of M33. Importantly, M31's COM is not fixed in space for our main results; rather, it is allowed to move in response to M33 \citep[see also][]{gomez15}. We elaborate on the consequences of a fixed M31 COM in Section \ref{sec:orbit_results}.

We note that \citet{corbelli14} and \citet{kam17} report even higher total masses for M33 between $4-5 \times 10^{11}\, M_{\odot}$, making our highest M33 mass model most favorable in the context of M33 literature mass estimates. We refer readers to \citet{corbelli24}, which carries out an in-depth investigation of the M33-M31 interaction assuming a total M33 mass in the range $3-5 \times 10^{11}\, M_{\odot}$.  

\subsubsection{Backwards Orbit Integration}

Following \citet{sohn20} and \citet{metz07}, we rotate the coordinates in Table~\ref{tab:coords} to the M31-centric frame using the rotation matrix provided in Equation~7 of \citet{sohn20}. This matrix transforms coordinates from Galactocentric $\rm (X,Y,Z)_{GC}$ to a system with M31 at its origin. The coordinates are also specific to the inclination and position angle of M31 \citep[see][]{metz07}. The M31 disk is aligned with the XY plane in this reference frame and the X-axis points away from the center of the MW (line-of-sight depth) such that the closer side has $-x$. The Y-axis is perpendicular to the X-axis along the disk plane, so $+y$ is approximately the vector from SW to NE in M31's disk plane.  

Using these methods combined with the 6D phase space coordinates in Table~\ref{tab:coords}, we will explore the plausible orbital histories for M33 in both the low and high mass M31 potentials. Orbits are integrated for the last 6 Gyr using a leapfrog algorithm. These will be referred to as \emph{direct orbital histories}. We account for uncertainties in the galaxies' 6D phase space information by calculating 1,000 unique orbits per each M31-M33 mass combination for both observational data sets, as in previous work \citep[e.g.,][]{sohn17, sohn20, patel20}. Each of these 1,000 orbits is initialized with a slightly different 6D phase space vector, corresponding to a sample drawn from a probability distribution representing the joint 1$\sigma$ uncertainties on M33 and M31's distance, line-of-sight velocity and proper motions in a Monte Carlo fashion.

\section{Orbital Histories of the M31-M33 System}
\label{sec:orbit_results}

A variety of M33 orbital histories have been proposed in the literature to date. M33's proper motion, and therefore its 3D velocity when combined with measurements of its line-of-sight velocity, has been known since \citet{brunthaler05}. However, in the absence of a measurement for M31's 3D velocity, which only became available in 2012 \citep{sohn12, vdm12ii}, preceding attempts to constrain M33's orbit aimed to reproduce its morphology \citep{mcconnachie09, putman09}. 

Earlier attempts to match M33's present-day morphology, which features warps in the outskirts of both its stellar and gaseous disks, proposed orbital solutions where M33 and M31 had a recent ($\sim$3 Gyr ago), close (50-100~kpc) tidal interaction \citep{mcconnachie09, putman09}. However, this orbit does not match the now-known motion of M33 relative to M31. 

More recently, using the available 6D phase space information for M31 and M33, \citetalias{patel17a} showed that M33 is only approaching its closest distance relative to M31 today and is statistically most likely to be in a first infall scenario if M31's total mass is $< 2 \times 10 ^{12} \, M_{\odot}$. Alternatively, in a more massive M31 potential ($\geq 2 \times 10 ^{12} \, M_{\odot}$), M33  completes a pericentric passage at $\sim$ 6 Gyr ago but only at a distance of $\sim$100~kpc \citep[see also][]{teppergarcia20}. With new M33 and M31 proper motions from \emph{Gaia} DR2, \citet{vdm19} found further evidence in support of the first infall scenario for M33, regardless of M31's mass. 

Other works \citep[e.g.,][]{semczuk18} have considered different values for M31's transverse motion \citep[e.g.,][]{salomon16} estimated via indirect methods such as through satellite kinematics, finding M33 could have completed a pericentric passage around M31 at a distance of 37~kpc at $\sim$ 3 Gyr ago. 
Newer direct measurements of M31's transverse motion have since been reported \citep{vdm19, salomon21, rusterucci24} with \emph{Gaia} data. Here, we take advantage of several recent literature measurements for M31's tangential velocity to explore two main orbital scenarios: one where M33 is on first infall into the halo of M31 like the LMC around the MW \citep[e.g.,][]{shaya13, patel17a}, and another where M33 passes around M31 several billion years ago at a large pericentric distance \citep[$\gtrsim150$~kpc; e.g.,][]{patel17a, teppergarcia20}.

\subsection{Orbital Histories with M31's Center of Mass Artificially Fixed}
\label{subsec:artificial}

Although the implications of assuming a moving COM in the MW have been studied extensively \citep[e.g.,][]{white83, bekki12, gomez15, perryman14, bailin05, bryan07, veraciro11}, analogous effects on the M31 system have yet to be quantified. Therefore, we first demonstrate how artificially fixing M31's COM in an inertial reference frame yields different solutions for M33's orbital history.

\begin{figure}[t]
    \centering
    \includegraphics[width=0.45\textwidth, trim=5mm 0mm 0mm 0mm]{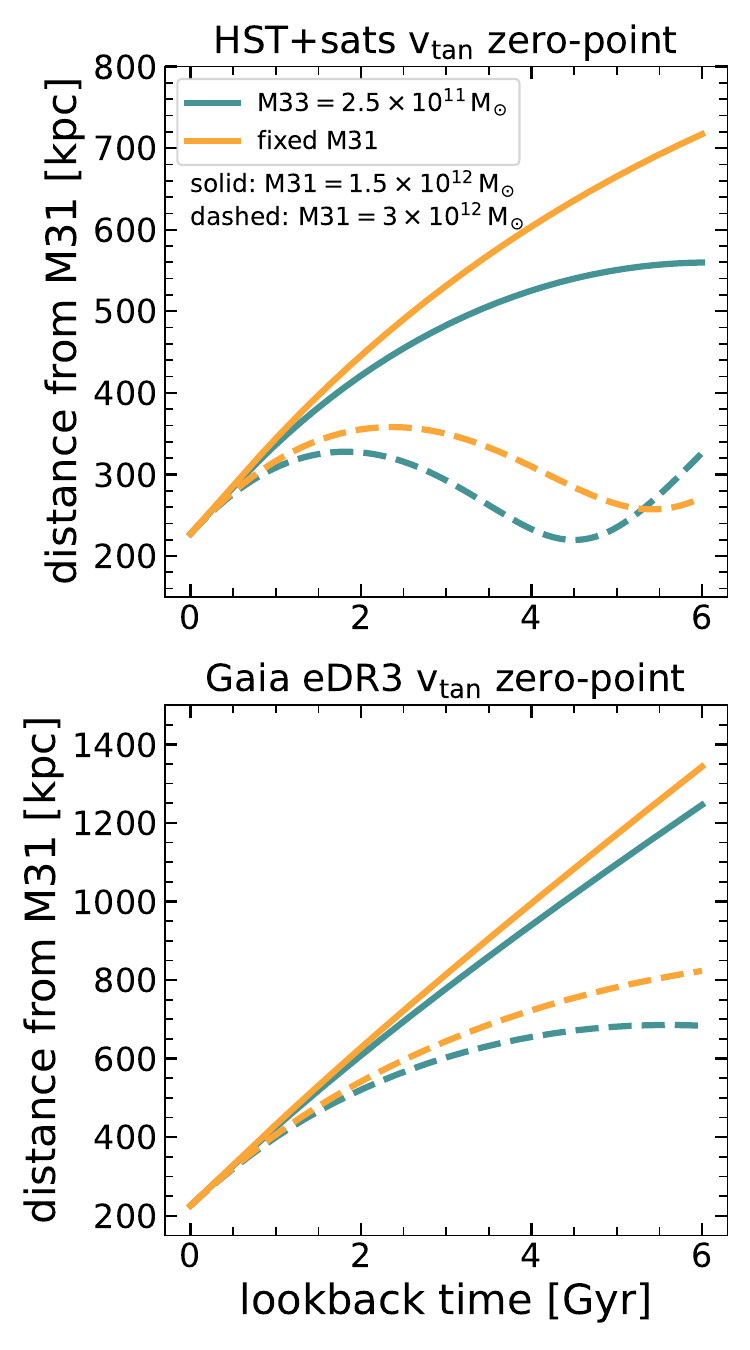}
    \caption{Orbital histories for the highest mass M33 model ($2.5\times10^{11}\,M_{\odot}$) in both a low (solid lines) and high mass (dashed lines) M31 potential. The top panel illustrates orbits using the HST+sats M31 tangential velocity, while the bottom panel uses the \emph{Gaia} eDR3 data. Blue-green lines represent 2-body orbits where M31's COM is free to move in response to M33, and the orange lines correspond to orbits when M31's COM is artificially fixed. Allowing M31's COM to move freely decreases the orbital period and distances along M33's orbital trajectory. }
    \label{fig:fixedvsmovingM31}
\end{figure}

Figure \ref{fig:fixedvsmovingM31} shows the orbital history of M33 using the highest mass M33 considered in this work ($2.5\times10^{11}\,M_{\odot}$) and both a low (solid lines) and high mass (dashed lines) M31 potential. Results are shown for both the HST+sats (top panel) and \emph{Gaia} eDR3 (bottom panel) M31 tangential velocity. Blue-green lines represent M33's orbit when M31's COM is allowed to move. The orange lines show M33's orbit when M31's COM is artificially fixed at the origin of the M31-centric inertial frame. When M31's COM is artificially fixed, all M33 orbits exhibit an increased orbital period and larger distances. Though freeing M31's COM does not yield an overall different orbital solution than the fixed scenario, the passage of M33 does indeed move the barycenter of the M31-M33 system. We quantify this further in subsequent sections.

As in previous work, we find M33 is either on first infall or has completed a pericentric passage $\approx$4.5-5.5~Gyr ago at distances of $\approx$200~kpc from M31 in both the moving and fixed M31 COM scenarios. In Section~\ref{sec:reflex}, we will further explore the response of M31's COM to the passage of M33.

\subsection{Orbital Histories with M31's Moving Center of Mass}

\begin{figure*}[t]
    \centering
    \includegraphics[width=0.95\textwidth, trim=12mm 0mm 0mm 0mm]{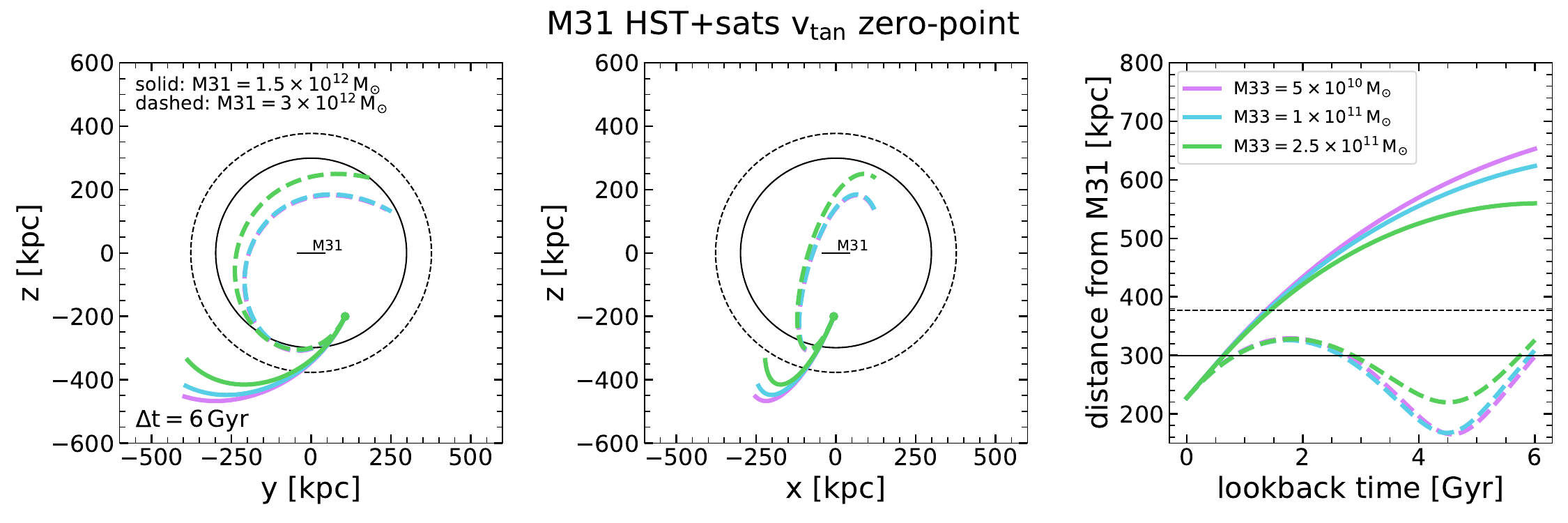}
    \includegraphics[width=0.95\textwidth, trim=12mm 5mm 0mm 0mm]{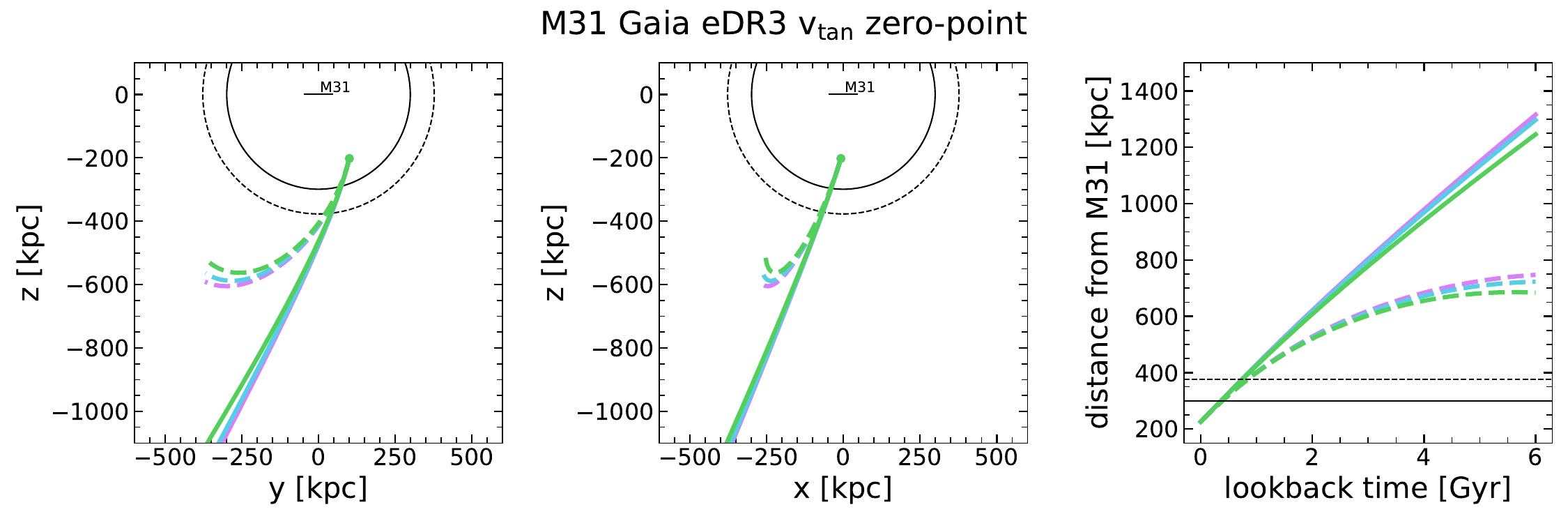}
    \caption{Top: The orbits of M33 over the last 6 Gyr using the HST tangential velocity for M31. Bottom: Orbits for M33 using the \emph{Gaia} eDR3 tangential velocity for M31. In both rows, results are shown for a low (solid lines) and high (dashed lines) mass M31, as well as three mass models for M33. The location of M33 is transformed to the M31-centric frame described in Section~\ref{subsec:orbit_methods}, where $z=0$ represents the projection of M31's disk plane, and the X-axis points away from the MW (line-of-sight depth). The Y-axis is such that $+y$ is approximately the vector from SW to NE in M31's disk plane. Dashed and solid black circles indicate the virial extent of each M31 model. The right-most panel shows the orbits of M33 as a function of lookback time.}
    \label{fig:m33orbits}
\end{figure*}

Figure \ref{fig:m33orbits} illustrates orbits for all M33 mass models using the HST+sats M31 tangential velocity (top panels) and the \emph{Gaia} eDR3 M31 tangential velocity (bottom panels). The first two columns show two projections of M33's orbit where $z=0$ is the projection of M31's disk. The cross-section panels (left and middle) also annotate the extent of the virial radius for both M31 potentials in the solid (low mass M31) and dashed (high mass M31) black circles. The right-most panel in Figure \ref{fig:m33orbits} shows the magnitude of relative distance between M33 and M31 as a function of lookback time. 

For the HST+sats orbits, all M33 mass models result in first infall in the low mass M31 potential. These results are in agreement with those reported in \citetalias{patel17a} where in a low mass M31, M33 was concluded to be on first infall having entered the halo of M31 $\sim$2 Gyr ago for the first time within a 6 Gyr integration period (see Appendix \ref{sec:appendixA}). This result is also consistent with the orbital history reported in \citet{shaya13} and predictions for the future of M33's orbit in \citet{vdm12iii}.

In the high mass M31 potential with the HST+sats M31 tangential velocity, M33 completes a pericentric passage at $>$4 Gyr ago at a range of distances ($>$ 160~kpc) proportional to the mass of M33. These results agree qualitatively with \citetalias{patel17a} but differ quantitatively. First, the high mass M31 in \citetalias{patel17a} had a virial mass of only $\rm M_{vir}= 2 \times 10^{12} \, M_{\odot}$, whereas, in this work, the high mass M31 potential is $\rm M_{vir}= 3 \times 10^{12} \, M_{\odot}$. Additionally, we use slightly different 6D phase space coordinates where M33 has a larger distance relative to M31 in this work than in the previous work. Noting these differences, Figure \ref{fig:m33orbits} shows a pericentric passage at 4.5~Gyr ago between 165-220~kpc, whereas \citetalias{patel17a} found a pericentric passage at about 6~Gyr ago at distances of 75-135~kpc.

As demonstrated in \citet{gomez15}, the influence of a massive satellite (like the LMC or M33) can substantially shift the orbital barycenter of a system, especially if the satellite is on first infall. The orbital barycenter of the M31-M33 system is shifted away from M31's present-day COM location by 7-32~kpc in the low mass M31 potential, regardless of the assumed M31 tangential velocity. For the high mass M31 potential, the barycenter shifts between 3-17~kpc relative to the COM of M31 at present-day. 

As the orbits shown in Figure \ref{fig:m33orbits} only indicate direct orbital histories for M33 (i.e., without factoring in the uncertainties on the 6D phase for both M33 and M31), Tables~\ref{tab:orbit_params1} and \ref{tab:orbit_params2} list summary statistics for orbital parameters in both the low and high mass M31 potentials. The tables indicate the fraction of 1,000 orbits with pericenters and apocenters. They also provide the distances and times at which these critical minima and maxima are reached. Uncertainties on orbital parameters represent the [15.9, 50, 84.1] percentiles. Values preceding the brackets are from the direct orbital histories presented in Figure~\ref{fig:m33orbits}. 

\begin{deluxetable*}{lcccccc}[h!]
\tablecaption{Orbital Parameters for M33 Relative to M31 with HST+sats $v_{\rm tan}$\label{tab:orbit_params1}}
\tablewidth{0pt}
\tablehead{
\colhead{M33 Mass}  &  \colhead{$\rm f_{peri}$} & \colhead{$\rm t_{peri}$} & \colhead{$\rm r_{peri}$} &  \colhead{$\rm f_{apo}$} & \colhead{$\rm t_{apo}$} & \colhead{$\rm r_{apo}$} \\
\multicolumn1c{($10^{10}\, M_{\odot})$} & \colhead{(\%)} &  \colhead{(Gyr)} & \colhead{(kpc)} &  \multicolumn1c{(\%)} & \multicolumn1c{(Gyr)}  & \multicolumn1c{(kpc)} 
}
\startdata
 \multicolumn{7}{c}{\textbf{Low Mass M31}}\\  \hline 
5 & 6.6 & $\cdots$ [3.03, 4.53, 5.60] & $\cdots$ [158, 200, 234] & 29.5 & $\cdots$ [2.15, 3.51, 5.07] & $\cdots$ [295, 377, 469] \\
10 & 7.7 & $\cdots$ [2.99, 4.60, 5.67] & $\cdots$ [154, 195, 236] & 32.2 & $\cdots$ [2.11, 3.42, 5.02] & $\cdots$ [295, 372, 477] \\
25 & 9.0 & $\cdots$ [2.57, 4.23, 5.69] & $\cdots$ [168, 213, 256] & 37.6 & $\cdots$ [2.30, 3.48, 4.98] & $\cdots$ [305, 389, 484] \\ \hline
 \multicolumn{7}{c}{\textbf{High Mass M31}}\\  \hline 
5 & 64.2 & 4.57 [2.87, 4.07, 5.32] & 165 [112, 150, 192] & 91.2 & 1.79 [0.96, 1.93, 3.28] & 329 [264, 338, 462] \\
10 & 65.6 & 4.48 [2.83, 4.01, 5.27] & 167 [113, 151, 195] & 92.1 & 1.74 [0.93, 1.88, 3.2] & 325 [263, 336, 459] \\
 25 & 65.7 & 4.5 [2.68, 3.91, 5.27] & 219 [138, 188, 246] & 93.2 & 1.82 [0.94, 2.03, 3.29] & 327 [263, 339, 470] \\ \hline
\enddata
\tablecomments{Orbital parameters for M33 in the low mass and high mass M31 potential with the HST+sats M31 $\rm v_{tan}$. Columns represent the fraction of 1,000 orbits where a pericentric (apocentric) passage is recovered, the lookback time at which these critical points occurred, and the M31-centric distance of M33 at that lookback time. The quoted uncertainties represent the [15.9, 50, 84.1] percentiles. Parameters with missing data are those instances where a pericentric/apocentric passage occurred $>$ 6~Gyr ago or will occur in the future. Values preceding the square brackets correspond to the direct orbital histories shown in Figure \ref{fig:m33orbits}.}
\end{deluxetable*}

\begin{deluxetable*}{lcccccc}[h!]
\tablecaption{Orbital Parameters for M33 Relative to M31 with \emph{Gaia} eDR3 $v_{\rm tan}$ \label{tab:orbit_params2}}
\tablewidth{0pt}
\tablehead{
\colhead{M33 Mass}  &  \colhead{$\rm f_{peri}$} & \colhead{$\rm t_{peri}$} & \colhead{$\rm r_{peri}$} &  \colhead{$\rm f_{apo}$} & \colhead{$\rm t_{apo}$} & \colhead{$\rm r_{apo}$} \\
\multicolumn1c{($10^{10}\, M_{\odot})$} & \colhead{(\%)} &  \colhead{(Gyr)} & \colhead{(kpc)} &  \multicolumn1c{(\%)} & \multicolumn1c{(Gyr)}  & \multicolumn1c{(kpc)} 
}
\startdata
 \multicolumn{7}{c}{\textbf{Low Mass M31}}\\  \hline
5 & 0.0 & $\cdots$ [0.0, 0.0, 0.0] & $\cdots$ [0, 0, 0] & 1.7 & $\cdots$ [3.39, 4.40, 5.76] & $\cdots$ [389, 452, 535] \\ 
10 & 0.0 & $\cdots$ [0.0, 0.0, 0.0] & $\cdots$ [0, 0, 0] & 1.8 & $\cdots$ [3.19, 4.49, 5.20] & $\cdots$ [379, 454, 522] \\
25 & 0.0 & $\cdots$ [0.0, 0.0, 0.0] & $\cdots$ [0, 0, 0] & 3.2 & $\cdots$ [3.26, 4.32, 5.69] & $\cdots$ [370, 475, 536] \\ \hline
 \multicolumn{7}{c}{\textbf{High Mass M31}}\\ \hline 
5 & 11.4 & $\cdots$ [4.16, 5.10, 5.74] & $\cdots$ [95, 151, 173] & 40.7 & $\cdots$ [2.18, 3.40, 4.96] & $\cdots$ [374, 492, 613] \\
10 & 12.4 & $\cdots$ [4.12, 5.08, 5.70] & $\cdots$ [97, 149, 176] & 43.2 & $\cdots$ [2.17, 3.35, 4.92] & $\cdots$ [372, 491, 618] \\
25 & 12.9 & $\cdots$ [4.09, 5.19, 5.73] & $\cdots$ [110, 184, 227] & 47.0 & 5.59 [2.31, 3.41, 4.99] & 685 [381, 494, 624] \\ \hline
\enddata
\tablecomments{Same as Table \ref{tab:orbit_params1} for the M31 \emph{Gaia} eDR3 $\rm v_{tan}$. }
\end{deluxetable*}

Table \ref{tab:orbit_params1} indicates the probability of M33 having a pericentric passage in a low mass M31 is only common for 7-9\% of orbits using the HST+sats M31 tangential velocity. However, as shown by the direct orbital histories, the distances achieved at pericenter range from 160-260~kpc. For the high mass M31 potential (bottom, Table \ref{tab:orbit_params1}), a pericentric passage is recovered for $\sim$65\% of orbits for all M33 mass models, however, the distances at pericenter are also large ranging from 110-250~kpc. As concluded in \citetalias{patel17a}, a pericentric passage more than a few billion years at distances $>$100~kpc, is inconsistent with a recent tidal interaction with M31 as proposed in \citet{mcconnachie09} and \citet{putman09}. This orbital solution is also different from that of \citet{teppergarcia20}, who found a pericentric passage at $\approx$6~Gyr ago but also at distances of only $\approx$50~kpc. At the 3$\sigma$ level, it is possible to find orbital solutions where M33 reaches a pericentric distance of 63-80 kpc from M31 in the last 1.6-1.8 Gyr in the high mass M31 potential, while the corresponding values are 110-126 kpc at 1.0-1.3 Gyr ago in the low mass M31 potential.

The bottom panels of Figure~\ref{fig:m33orbits} illustrate orbits using the \emph{Gaia} eDR3 M31 tangential velocity. These panels show that M33 is on first infall, only reaching its closest approach to M31 today for both the low (solid lines) and high (dashed lines) mass M31 models. These results are consistent with those reported in \citet{vdm19}, which showed that \emph{Gaia} DR2-based proper motions for M33 and M31 strengthened the arguments of \citetalias{patel17a} that M33 is on first infall into the halo of M31. Furthermore, Table~\ref{tab:orbit_params2} also confirms that it is statistically rare to recover orbits where a pericentric passage has occurred. Zero orbits show evidence of a pericenter for the low mass M31. Only 11-13\% of orbits in the high mass M31 potential recover a pericenter at distances of $\sim$100-230~kpc between 4-6 Gyr ago (see Table \ref{tab:orbit_params2}) on average, still much larger than those required by previous studies that include a recent, tidal encounter \citep{mcconnachie09, putman09, semczuk18}. At the 3$\sigma$ level, pericenter distances of 55-65 kpc relative to M31 are possible 2.8-3.0 Gyr ago in the high mass M31 potential.

Given the distances and timescales recovered in this work, it is unlikely that tidally induced warps in M33 by M31 would survive until today \citep[see][]{semczuk20}. Both the first infall results and a pericentric passage at $>$4~Gyr ago are also consistent with the lack of a truncated gas disk in M33 \citep{putman09}. Additionally, both \citet{dobbs18} and \citet{corbelli24} provide simulation-based arguments for alternative explanations to M33's warps, namely \citet{dobbs18} suggests stellar feedback and gravitational instabilities while \citet{corbelli24} lend the explanation of M33's warps to gas accretion from cosmic filaments and outer disk misalignment in late-type galaxies. 

Additionally, \citet{corbelli24}\ accounts for a higher mass M33 and also finds a low probability of a past M33-M31 encounter with the \emph{Gaia} eDR3 M31 tangential velocity. In instances where an interaction is possible, their results agree with those presented here and in \citetalias{patel17a}, which suggest the minimum distance between M33 and M31 in these encounters is $\gtrsim$ 100~kpc. Thus, the literature and our results together provide plausible explanations for a first infall scenario and an explanation for M33's present morphology. 

\section{Results: M31's Center of Mass Evolution Owing to M33}
\label{sec:reflex}
In this section, we quantify the changes in M31's COM phase space coordinates due to M33's passage. We explore all M33-M31 mass combinations, yielding mass ratios of 1:6, 1:15, and 1:30 in the low mass M31 potential and 1:12, 1:30, and 1:60 in the high mass M31 potential. 

\subsection{Magnitude of M31's COM Position and Velocity Evolution}
\begin{figure*}[t]
    \centering
    \includegraphics[width=0.8\textwidth, trim=0mm 5mm 0mm 10mm]{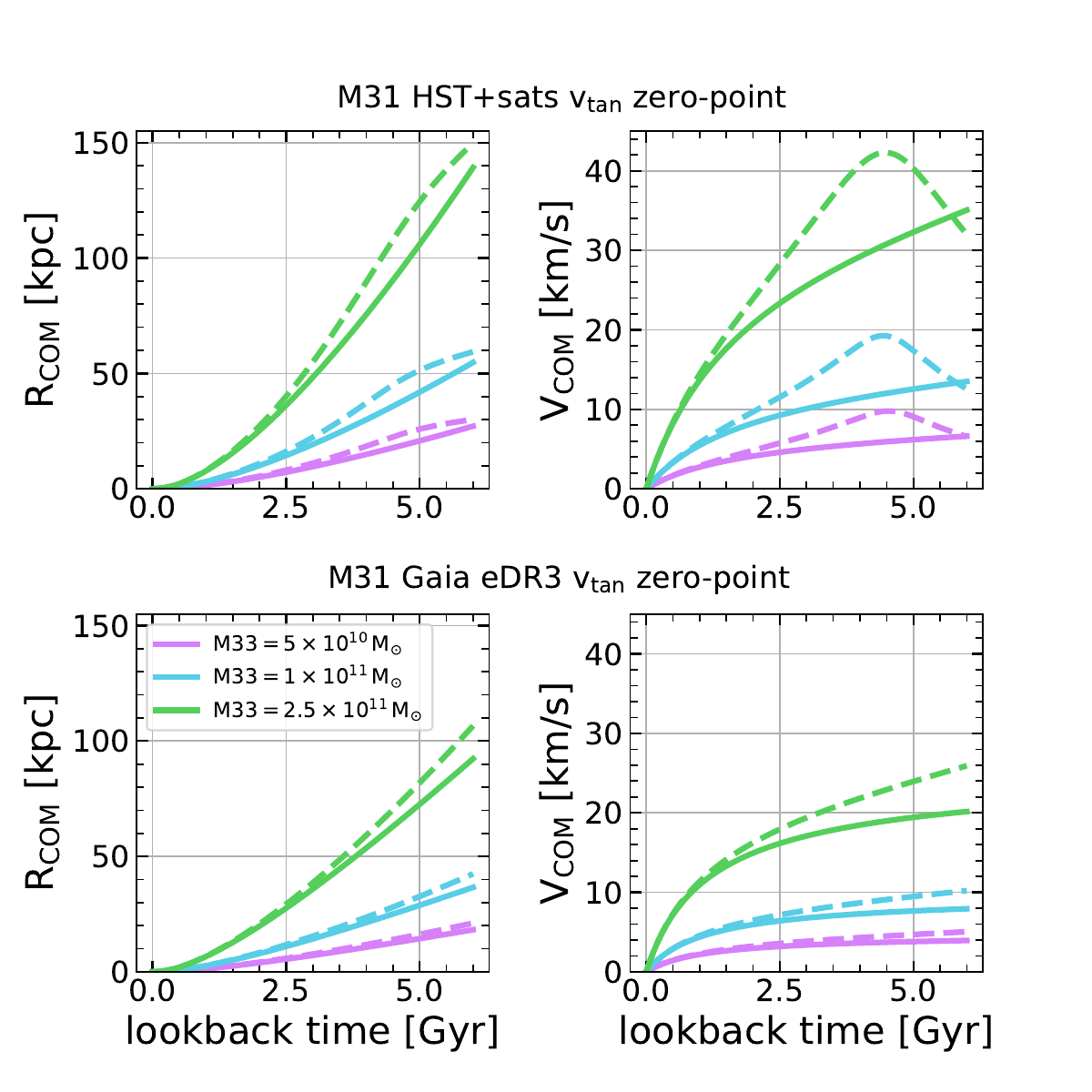}
    \caption{Time evolution of M31's COM position and velocity relative to its present position and velocity. M31's COM begins at (0,0,0)~kpc and (0,0,0) km s$^{-1}$, and it is allowed to move in response to M33's gravitational influence as orbits are integrated backward in time. The left panels show the magnitude of M31's COM position, and the right panel shows the corresponding COM velocity for three different M33 masses. \textbf{Top}: Results for the HST+sats tangential velocity direct M33 orbits (see Figure \ref{fig:m33orbits}, right panels). Solid and dashed lines correspond to the low and high mass M31 ($3 \times 10^{12} \, M_{\odot}$), respectively. \textbf{Bottom}: Same as top panels but for M33 direct orbits derived using the $Gaia$ eDR3 M31 tangential velocity. Since M33 does not pass through pericenter in these orbits, M31's COM position and velocity steadily decrease approaching present-day. }
    \label{fig:m31COM}
\end{figure*}

To determine the gravitational influence of M33 on M31's COM, we first examine the shift of M31's COM position from the origin of the M31-centric reference frame [(0,0,0)~kpc] and the corresponding evolution of M31's COM velocity. Figure~\ref{fig:m31COM} illustrates the evolution of M31's COM position (\rcom{}, left panel) and velocity (\vcom{}, right panel) corresponding to the M33 orbits shown in Figure \ref{fig:m33orbits}. The top panels refer to the HST+sats M31 tangential velocity, and the bottom panels refer to the \emph{Gaia} eDR3 orbit results. Similar to Figure \ref{fig:m33orbits}, Figure \ref{fig:m31COM} shows results for both the low (solid lines) and high mass (dashed lines) M31 potentials.

As expected, the top panels of Figure \ref{fig:m31COM} show M31's response to M33 increases with the mass of M33, or as the mass ratio between the two galaxies increases. Over the last 6 Gyr, M33 causes \rcom{} to move by 30-140~kpc in the low mass M31 potential. The corresponding \vcom{} ranges from 6-35 km s$^{-1}$. For the high mass M31, M31's COM position ranges from 30-150~kpc, and its COM velocity ranges from 6-42 km s$^{-1}$. Note that the COM velocity peaks where M33 passes through the pericenter at $\sim$4.5~Gyr ago. 

The bottom panels in Figure \ref{fig:m31COM} correspond to the M33 orbits resulting from the $Gaia$ eDR3 tangential velocity for M31. In a first infall orbit, M31's COM position and velocity reach a smaller magnitude than with the HST+sats M33 orbits, allowing M33 to get closer to M31. Regardless, for the low mass M31, M31's \rcom{} ranges from 18-92~kpc, and the corresponding \vcom{} is 4-20 km s$^{-1}$. In the high mass M31 potentials, these effects increase to \rcom{}=21-108~kpc and \vcom{}=5-26 km s$^{-1}$.

The combination of M31/M33 masses where M31's \rcom{} and \vcom{} are most affected is for the highest M33 mass ($2.5 \times 10^{11}\, M_{\odot}$) and the high M31 mass. Examining the 1,000 orbits computed in Section \ref{subsec:orbit_methods} for this mass combination, the COM position when M33 passes through pericenter is \rcom{} = $92.6 ^{+34.8}_{-32.9}$~kpc and the range of COM velocities when M33 passes through pericenter is constrained to \vcom{} = $43 ^{+5.9}_{-4.7}$ km s$^{-1}$ where the uncertainties give the [15.9, 84.1] percentiles around the median of the distribution. The timing of M33's pericenter for this mass combination is $t_{\rm peri}=4.5^{+0.77}_{-1.82}$ Gyr ago, as indicated in Table \ref{tab:orbit_params2}. Appendix B includes additional figures quantifying the uncertainties on the magnitude of M31's COM position and velocity as a function of time.

Unlike in the MW-LMC interaction, where 80\% of the displacement and reflex motion imparted by the LMC on the MW's COM occurs in the last 1 Gyr, the evolution of the M31-M33 system is less drastic. Since M33 does not reach as deep into the potential well of M31 as the LMC does in the MW's halo, 50\% of M31's COM position and velocity evolution occurs over the last $\sim$ 4 Gyr, independent of M31's mass and choice of tangential velocity. Despite the slower interaction between M33 and M31, we conclude M33 does impart a non-negligible offset to the COM location and motion of M31, and these phenomena must be taken into account in dynamical studies of the M31 system. In Section~\ref{sec:discussion}, we discuss implications for the orbits and spatial distribution of other M31 satellites.

Interestingly, \citet{salomon23} identifies a host-satellite pair in the HESTIA simulations that is analogous to M33, specifically a massive satellite analog of approximately 10\% of its host galaxy's mass that passes around the host at 130~kpc. This induces a velocity offset of 66 km s$^{-1}$ between the host disk's COM and the COM inferred using just the most massive satellite, comparable to our results for M33 in a high mass M31 with the HST+sats data set. More generally, larger offsets in both position and velocity are reported between the disk COM and the satellites compared to the disk COM and the halo COM. As such, we speculate this offset biases existing estimates of M31's tangential velocity inferred from only satellite kinematics, which assume a COM position that may not accurately represent the real COM of the disk \citep[e.g.,][]{vdmG08, salomon16}. This is beyond the scope of this work and will be the subject of future work.

\begin{figure*}
    \centering
    \includegraphics[width=0.85\textwidth, trim=0mm 0mm 0mm 0mm]{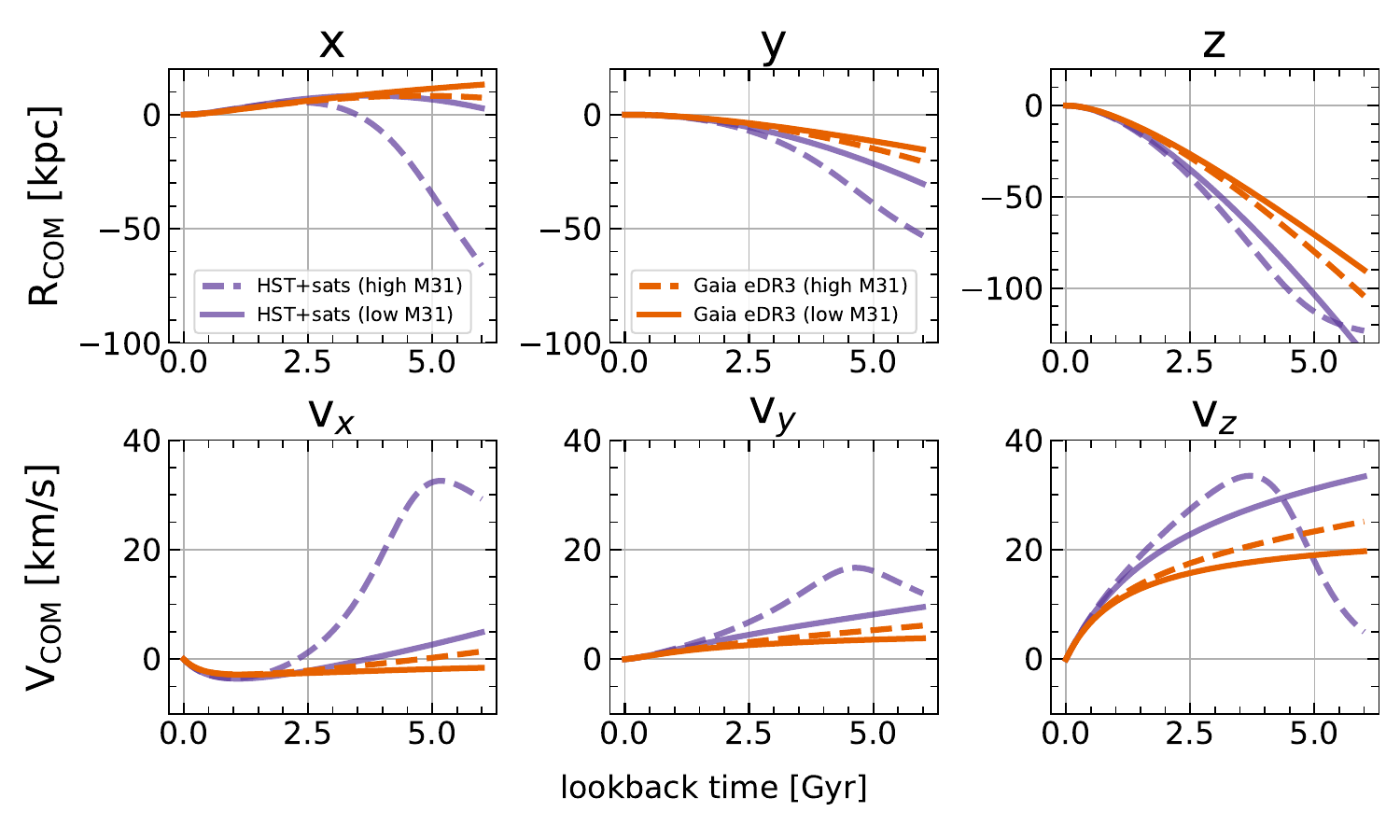}
    \caption{Components of M31's COM position (\rcom{}, top panels) and velocity (\vcom{}, bottom panels). All panels show the results for the highest mass M33 ($2.5 \times 10^{11} \, M_{\odot}$) in the low mass (solid lines) and high mass (dashed lines) M31 potentials. Orange and purple lines correspond to results using the $Gaia$ eDR3 and HST+sats tangential velocity orbits, respectively. The $z$ and $v_z$ components dominate the evolution of M31's \rcom{} and \vcom{} when M33 is on a first infall orbit (all \emph{Gaia} eDR3 orbits and the high mass HST+sats orbits). However, for the HST+sats results in the low mass M31 potential, the $x$ and $v_x$ components contribute equivalently to $z$ and $v_z$ in both \rcom{} and \vcom{}, albeit at different times. The dominant motion in the z-direction (orthogonal to M31's disk plane) is promising for potential observational signals of M33's impact on M31's COM (see Section \ref{subsec:observables}).}
    \label{fig:m31COM_components}
\end{figure*}

\subsection{Direction of Maximal COM Position and Velocity Evolution}

In addition to quantifying the magnitude of M31's evolving COM position and velocity, we investigate the direction where the quantities are at maximum. Figure \ref{fig:m31COM_components} decomposes the COM position and velocity of M31. The top panels show the COM position in $x, y, z$ in the highest mass M33 model as a function of time. The bottom panels in Figure \ref{fig:m31COM_components} show the evolution of the COM velocity by component. Orange curves denote M33's orbit using the $Gaia$ eDR3 tangential velocity, while the blue curves correspond to the HST+sats tangential velocity. As in previous figures, solid (dashed) lines correspond to the low (high) mass M31 potential. 

For the $Gaia$ eDR3 orbits, the $z$ and $v_z$ components of COM position and velocity are always dominant. We remind readers that in our adopted coordinate system, the Z-axis is normal to M31's disk plane. This is expected as the motion of M33 is primarily in the $z$ direction in the first infall scenario illustrated in the bottom row of Figure \ref{fig:m33orbits}. For the HST+sats results, $z$ and $v_z$ also exhibit the most evolution in the low mass M31 potential, where M33 is also in a first infall scenario. However, for the HST+sats results in the high mass M31 potential, the $v_x$ and $v_z$ components reach equivalent magnitudes at maximum, albeit at different times, similar to the $x$ and $z$ components. 

In Section~\ref{subsec:distribution}, we will determine whether M31's response to M33's passage also plays a role in the distribution and orbits of other M31 satellites. Additionally, we predict the observable magnitude of M31's COM motion in response to M33 and discuss the feasibility of observing this motion with existing and upcoming surveys of M31's halo.

\section{Discussion}
\label{sec:discussion}
Thus far, we have established M33 does indeed induce a dynamical response from M31, namely M31's COM is displaced by tens of kiloparsecs, and its velocity is offset by up to a few tens of kilometers per second. While M33 is the most massive satellite orbiting M31, M31 also hosts another $\sim$35 satellite galaxies \citep[e.g.,][]{mcconnachie18}. For ten of these galaxies, 6D phase space information is now available, allowing us to quantify whether M31's response to the passage of M33 has implications for the orbits and spatial distribution of $\sim$30\% of M31's satellite system. In this section, we provide predictions for where M31's velocity offset between the motion of the outer halo relative to the inner halo, or reflex motion, is maximized. We also discuss the feasibility of measuring the reflex motion using M31 halo stars. A measurement of reflex motion is otherwise known as a travel velocity and has been measured for the MW-LMC system \citep{erkal21, petersen21, chandra24}.

Throughout the discussion, we will default to the HST+sats M31 tangential velocity as this allows for two possible M33 orbital histories: a first infall orbit for the low mass M31 potential and an orbit including a wide pericentric passage ($\rm r_{peri} \sim$ 200~kpc) in the high mass M31 potential (see Figure \ref{fig:m33orbits}). This combination of parameters also maximizes M31's response to the passage of M33, providing a cleaner test of COM velocity offsets due to massive accretions versus COM velocity offsets due to passive evolution \citep[see][]{salomon23}.

\subsection{Consequences of M31's Response to M33 on M31 Satellite Orbits}
\label{subsec:consequences}

\begin{figure*}[t]
    \centering
    \includegraphics[width=0.99\textwidth, trim=0mm 0mm 0mm 0mm]{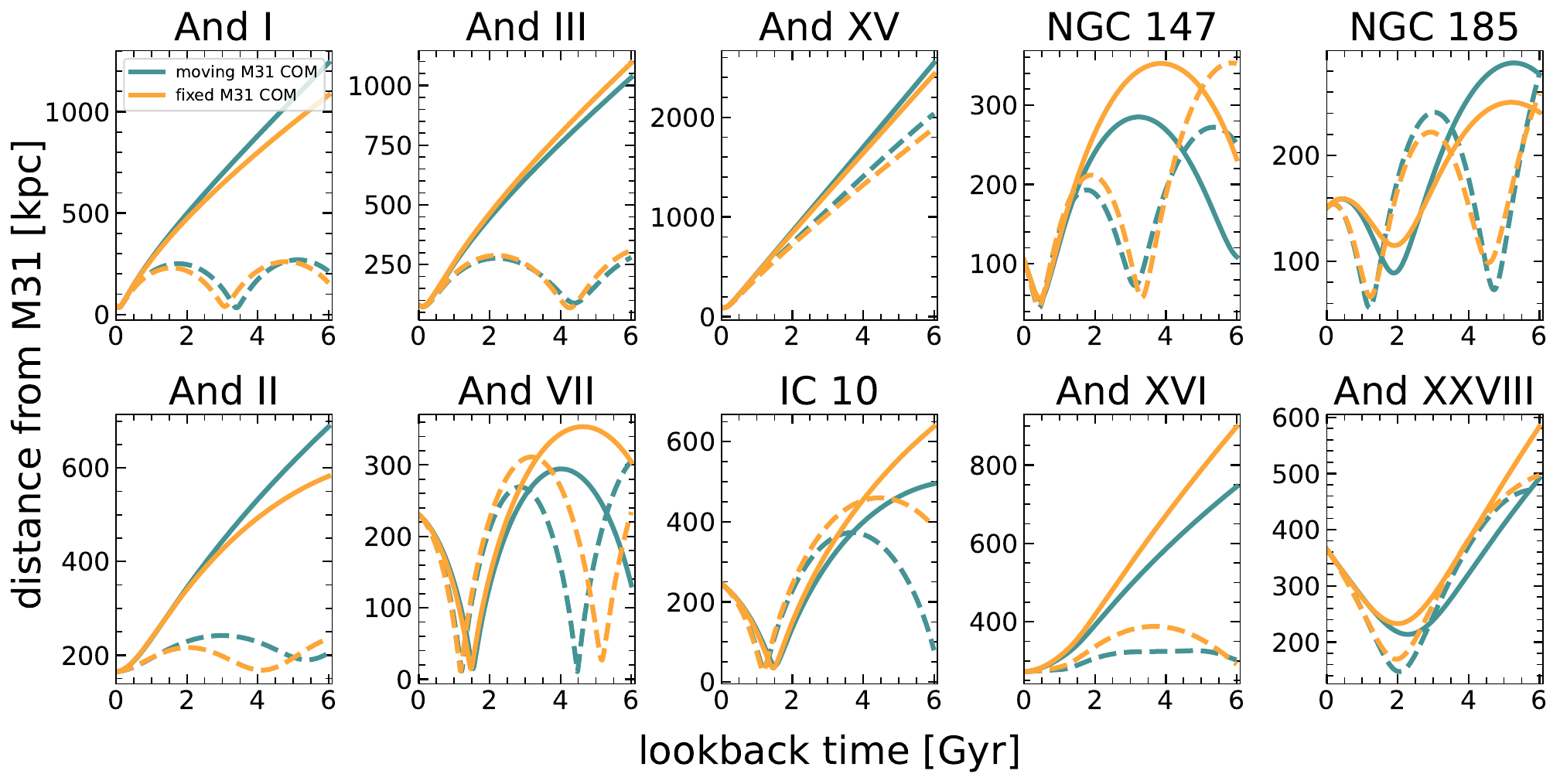}
    \caption{3-body orbits for ten M31 dwarfs in the low (solid lines) and high mass (dashed lines) M31 potential, including the influence of M33. All orbits adopt the HST+sats M31 tangential velocity zero-point. Satellites are ordered by increasing distance from M31. Blue-green lines indicate orbits where M31 is free to move in response to the other two bodies, while orange lines show results for an artificially fixed M31. M31's COM response to M33, combined with the gravitational influence of M33 on the dwarfs themselves, impacts the orbits of all ten satellites.}
    \label{fig:dwarf_orbits}
\end{figure*}

It is now well-known that massive satellite galaxies like the LMC and M33 can significantly influence the orbits of other satellite galaxies as they simultaneously orbit around their host galaxies. Backward orbit integration shows the influence of the LMC on lower mass dwarfs around the MW has a variety of effects: some satellites show dynamical association with the LMC, implying a group infall scenario; some satellites experience a swift fly-by encounter with the LMC, and others may even be captured by the LMC during its infall into the MW's halo \citep[e.g.,][]{patel20, battaglia22}.

Given the equivalent mass ratios between the MW-LMC and M31-M31 systems, here, we investigate two phenomena: how artificially fixing M31's COM affects the orbits of low mass M31 satellites and how the gravitational influence of M33 affects the orbits of other M31 satellites.

The ten M31 satellites considered in this analysis include NGC 147 and NGC 185, whose HST proper motions were first published in \citet{sohn20}, and IC 10, whose proper motion was measured with water masers in \citet{brunthaler07}. Additionally, we consider seven other M31 dwarf satellites: And~I, And~II, And~III, And~VII, And~XV, And~XVI, and And~XXVIII. These seven dwarfs have preliminary proper motions from HST GO-16273 and will be published in upcoming work (Sohn, Patel, et al., in prep.). For all galaxies, we adopt the distances provided in \citet{savino22}. 

We first calculate 2-body orbits between each satellite and M31 following the methodology outlined in Section \ref{subsec:orbit_methods}. The only modification is the choice of dynamical friction prescription. We adopt the prescription used in \citet{sohn20}, which is more well-tuned for satellite galaxies with mass ratios $<$ 1:10 \citep[see also][]{hashimoto03}. All ten satellite galaxies are modeled as Plummer spheres. Halo masses for NGC 147, NGC 185, and IC 10 are derived via the \citet{moster13} abundance matching relation: $5\times10^{10}\,M_{\odot}, 4.5\times10^{10}\,M_{\odot}$, and $3.7\times10^{10}\,M_{\odot}$. The corresponding scale lengths, computed by fitting a Plummer profile to estimated dynamical masses in the literature, are 5.0~kpc, 3.0~kpc, and 3.3~kpc \citep[see][]{sohn20}. The seven less massive dwarfs are all modeled as Plummer spheres with $M_{halo}=1\times10^{10}\,M_{\odot}$ and a scale length of 2~kpc.

For each satellite, two versions of 2-body orbits are computed, one where M31 is free to move in response to the satellites and another where M31 is artificially fixed in the inertial reference frame (as demonstrated in Fig.~\ref{fig:fixedvsmovingM31} for M33). Comparing these results, we find no difference in the orbits with a moving M31 COM, except for NGC~147 and NGC~185, the two most massive satellites of the ten considered, which can have orbits that differ in distance by up to 10\% at 6 Gyr ago. Allowing M31 to respond to their gravitational influence trends similarly to the results for M33 in Section~\ref{subsec:artificial} -- the orbital periods decrease by $\sim$5\%. Note that these conclusions are specific to the magnitude of the relative distance between M31 and the satellites and, therefore, do not account for any differences that arise in the components of position as a function of time.

Next, we consider how the passage of M33 influences these dwarfs. We calculate 3-body orbits for each dwarf using the highest mass M33 model and generate orbits where M31 is free to respond to other galaxies and where M31's COM is artificially fixed. The results of this exercise are illustrated in Figure \ref{fig:dwarf_orbits}. Here, blue-green lines indicate the orbits where M31's COM is free to move, and orange lines correspond to an artificially fixed M31 COM. Solid (dashed) lines represent the low (high) mass M31 potential results. Figure \ref{fig:dwarf_orbits} clearly shows M31's COM response to M33, combined with the gravitational influence of M33 on the dwarfs themselves, alters the orbits of all ten satellites. 

The dynamical impact of M33 on the orbits of the dwarfs increases with lookback time. NGC 147 and IC 10 are most significantly perturbed such that the relative distances to M31 at 6 Gyr ago differ by 150-300~kpc\footnote{This distance range refers to the magnitude of the vector resulting from subtracting i.e., NGC 147's position vector in a moving M31 COM  from that calculated in a fixed M31 COM potential at a fixed M31 mass.}. For NGC 147, IC 10, and And~VII, a moving M31 COM causes a shorter orbital period, regardless of M31's mass, similar to the results for M33 in Figure \ref{fig:fixedvsmovingM31}. And~XVI is affected to a similar degree as And~VII. For the remaining six dwarfs, M33's impact is still non-negligible. The relative distance between the dwarfs and M31 at 6 Gyr differs by $\sim$100~kpc, on average. Furthermore, the impact of M33 varies on a satellite-to-satellite basis, as demonstrated in \citet{patel20} for the inclusion of the LMC in MW satellite orbits. For some dwarfs, it leads to a shorter orbital period, as mentioned above, while for others, the orbital period may remain the same, but the pericentric and apocentric distances or timing changes. This is the case for NGC~185 in the low mass M31 where NGC 185's pericenter at $\sim$2 Gyr ago is $\sim$90~kpc in a moving M31 COM, but it increases to $\sim$120 when M31's COM is fixed. Similarly, NGC 185's apocenter at $\sim$5 Gyr ago differs by $\sim$40~kpc. 

\begin{figure}[h]
    \centering
    \includegraphics[width=0.48\textwidth, trim=12mm 0mm 0mm 0mm]{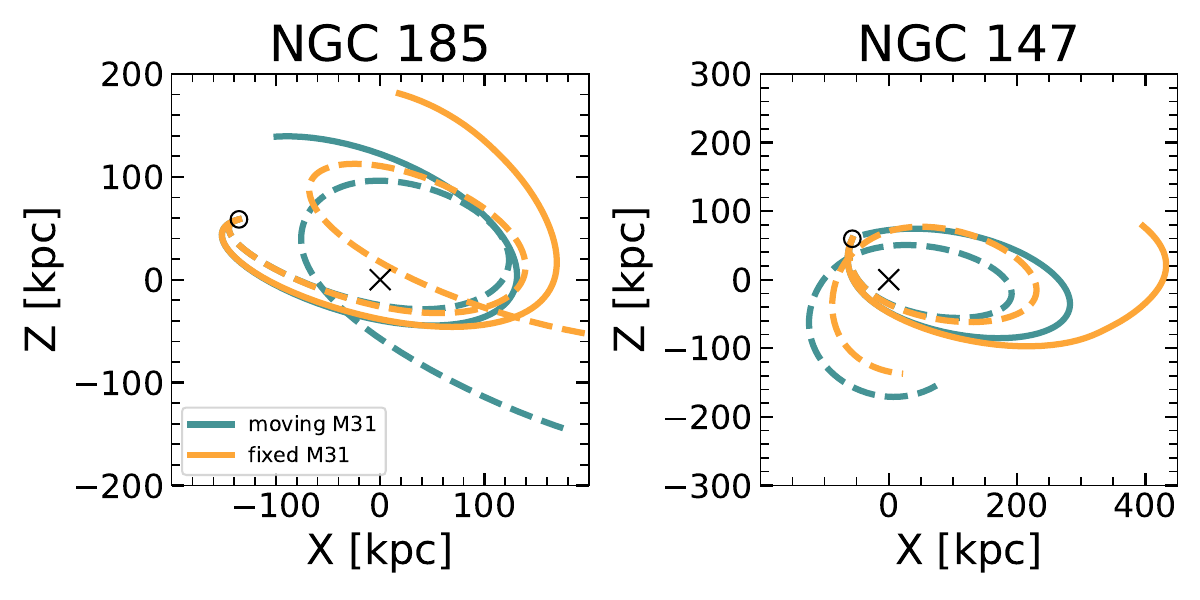}
    \caption{Orbits of NGC 147 and NGC 185, including the influence of M33 in the M31-centric XZ plane using the HST+sats M31 tangential velocity. Open circles are the positions of the satellite galaxies at present, while ``X" marks the location of M31 at present-day. Blue lines correspond to a moving M31 and orange lines show the orbits when M31's center of mass is artificially fixed. Solid (dashed) lines illustrate orbits in the low (high) mass M31 potential.}
    \label{fig:ngc_XZ}
\end{figure}

Figure \ref{fig:ngc_XZ} gives a sense of how these orbital perturbations manifest along different axes of direction. Figure \ref{fig:ngc_XZ} presents an alternative view of the joint impact of M33 and M31's response to M33 on the orbits of NGC 185 (left) and NGC 147 (right). This figure shows the $XZ$ orbital plane for the same data shown in Figure \ref{fig:dwarf_orbits}. It is clear that artificially fixing M31's COM significantly changes the predicted location of these galaxies at 6 Gyr~ago. For instance, in a low mass M31, NGC 185 is at ($x$, $z$) = -99, 139~kpc, whereas a fixed M31 yields ($x$, $z$) = 18, 181~kpc. 

We conclude M33's gravitational influence and the subsequent response of M31's COM to M33 is significant and must be taken into account when studying the accretion and evolutionary histories of all M31 satellites and other halo substructures. These phenomena are particularly relevant when studying the spatial distribution of the M31 satellite system, which is explored in Section \ref{subsec:distribution}.

We have not yet quantified the gravitational interplay between M33 on M32, a bright dwarf elliptical near M31's disk. Upcoming work (Patel et al., in prep.) will explore the possible interaction scenarios of the M31-M32-M33 system. Given its mass, proximity to M31, and yet-to-be-determined connection to the Giant Southern Stream, M32 may further amplify the evolution of M31's COM position and velocity. However, the orientation of M32's orbital plane relative to that of M33 is critical to whether the evolution is mitigated or becomes more prominent. While a lesser effect, NGC 147, NGC 185, and IC 10 are also likely to impart their own perturbations to M31's COM position and velocity. In fact, \citet{salomon23} attributes the offset in COM location and velocity between satellites and their host's disk to the few most massive satellites, warranting a further examination of M31's brightest dwarfs' (i.e., M32, NGC 185, NGC 147, IC 10) dynamical impact. Assessing the combined gravitational influence of the bright satellite galaxies on M31's COM will be especially necessary when more precise proper motions for M31 itself (HST GO-15658, PI: S.T. Sohn) are available.

\subsection{M33's Influence on the Spatial Distribution of M31 Satellites}
\label{subsec:distribution}

Using precise distances to M31 and its dwarf galaxies, \citet{savino22} showed that 80\% of M31 satellites lie in one hemisphere, the hemisphere aligned with the direction of the MW. Though this asymmetry was previously identified \citep[e.g.,][]{mcconnachie06,conn13}, the nature of this lopsided distribution is still unknown. Some theories for the origin of this lopsidedness include the recent accretion of many of the M31 dwarfs such that they have not yet had time to phase mix \citep{mcconnachie06}. Though Figure \ref{fig:dwarf_orbits} only includes ten M31 satellites and does not account for uncertainties in the measured 6D phase space information, only half of these satellites appear to have been accreted onto M31 recently. 

In the MW's halo, it has been shown that the passage of a massive satellite, particularly one that has had a recent, close pericenter like the LMC can enhance the phase space clustering of satellite galaxies \citep[e.g.,][]{gc21, gc24}. Here, we aim to determine whether the passage of M33 has had any analogous connection to the observed distribution of M31 satellites. 

\begin{figure*}[t]
    \centering
\includegraphics[width=0.98\textwidth]{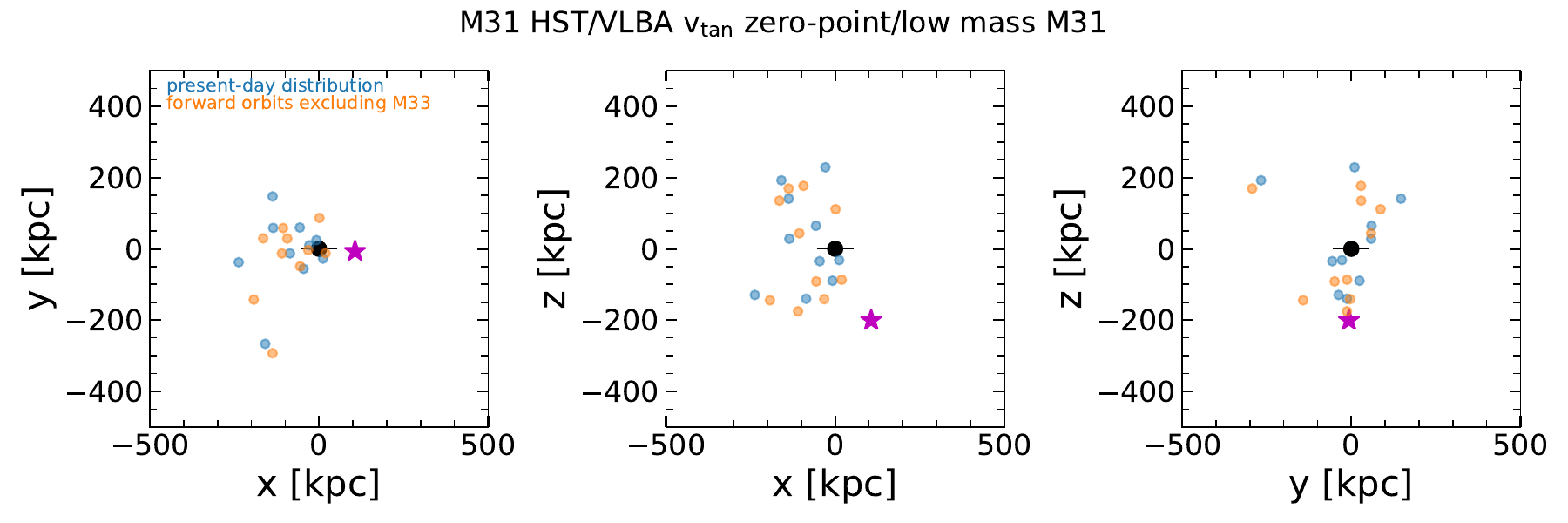}
\includegraphics[width=0.98\textwidth]{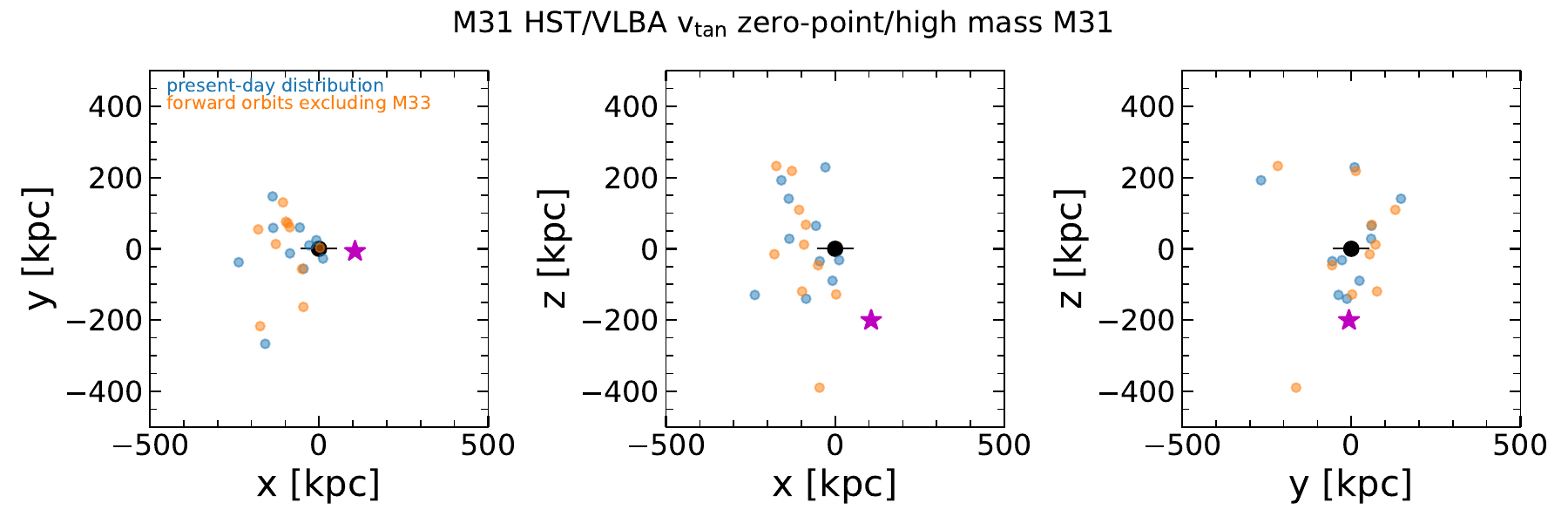}
    \caption{The spatial distribution of ten M31 dwarfs as observed today (blue points). The orange points show the spatial distribution by integrating backward until 6 Gyr ago in a 3-body scenario (M31+M33+dwarf), then using the coordinates at 6 Gyr to integrate 2-body orbits (M31+dwarf) forward to present day. The top (bottom) row corresponds to the low mass (high mass) M31 potentials. The magenta star represents the location of M33. The distribution of orange points shows that while M33 does influence the present-day location of M31 dwarf galaxies, it cannot explain the lopsided clustering of M31 satellites observed today.}
    \label{fig:spatial_distr}
\end{figure*}

The blue points in the top and bottom panels of Figure \ref{fig:spatial_distr} show the observed spatial distribution of the ten M31 satellites introduced in Section \ref{subsec:consequences}. Each panel corresponds to a cross-section in the M31-centric coordinate system\footnote{Our coordinate system is similar to the one used in \citet{savino22}, except for the direction of the y-axis.}. Though only ten satellites are shown, the asymmetry towards the $-x$ direction (the line-of-sight direction toward the MW) is evident.

To determine if M33 influences the distribution of these dwarfs, we use the 3-body orbits computed in Section \ref{subsec:consequences} to find the 6D phase space coordinates of every dwarf at 6 Gyr ago. With these past position and velocity vectors, we then integrate the 2-body orbit forward in time until today, excluding the influence of M33. The orange points in the top and bottom row of Figure \ref{fig:spatial_distr} correspond to the present-day spatial distribution resulting from this exercise. The top row is for the low mass M31 potential, and the bottom row uses the high mass M31 potential. Note that this exercise implicitly accounts for the influence of M33 on M31's COM.

The orange points in both rows show that while the distribution does shift when the influence of M33 is removed in the forward orbits, the asymmetric clustering is still apparent. Therefore, we conclude M33 is not driving the lopsidedness of the M31 satellite system, however, it may play a role along with other dynamical effects. The primary source of M31's lopsided satellite distribution, therefore, remains unknown. Importantly, this exercise only includes $\sim$30\% of M31's satellite system and may not be representative of the full M31 satellite system. 

Furthermore, the lopsided distribution is unlikely to be affected by the presence of the MW \citep{Santos-Santos24anisotropies} or the
result of a major merger $\geq 2-5$~Gyr ago \citep{kanehisa23}. The only way to study the evolution and possible origin of such lopsided distribution is via orbit reconstruction with proper motion measurements of M31 satellites. These measurements will allow us to trace the evolution of the lopsided distribution and should be revisited when 6D phase space information is available for the remaining M31 dwarfs.

\subsection{Predictions for Detecting M31's Travel Velocity Owing to M33}
\label{subsec:observables}
\begin{deluxetable}{cc|cc}
\tablecaption{Predictions for Travel Velocity in M31 Halo Stars\label{tab:prediction}}
\tablewidth{0.8\textwidth}
\tablehead{
\colhead{Distance}  &  \colhead{$\rm t_{dyn}$} & \colhead{$\rm  \Delta v_{rad, hel}$} & \colhead{$\rm \Delta v_{tan, hel}$}  \\
\colhead{(kpc)} & \colhead{(Gyr)} &  \colhead{(km s$^{-1}$)} & \colhead{(km s$^{-1}$)} 
}
\startdata
 \multicolumn{4}{c}{Low Mass M31}\\  \hline
100 & 1.6 & $\pm$8 & -13 \\
125 & 2.18 & $\pm$11 & -13 \\
150 & 2.82 & $\pm$16 & -12 \\\hline
 \multicolumn{4}{c}{High Mass M31}\\ \hline 
100 & 1.30 & $\pm$7 & -3 \\
125 & 1.75 & $\pm$10 & -3 \\
150 & 2.25 & $\pm$15 & -2\\\hline
\enddata
\tablecomments{Predictions for the observable shift in the radial and tangential velocities of stars in M31's halo relative to the motion of the disk (assumed to be M31's systemic velocity, $v_{\rm sys, M31} = 301$ km s$^{-1}$). Values are listed for a range of distances in M31's halo. Radial and tangential velocities are given in the heliocentric reference frame. Values are calculated for the highest mass M33 model and should be taken as upper limits. Spectroscopy of M31 halo stars, particularly RGBs, at these distances are most likely to contain imprints of M33's impact on M31's COM motion if the halo is indeed offset from the disk.}
\end{deluxetable}

In a live N-body model of M31 and M33's interaction history that captures the M33 orbits presented in this work, signatures of M31's COM displacement and reflex motion would be retained in the history of particles in M31's outer halo \citep[as is done in][for the MW-LMC system]{gc21}. However, our orbital models represent M31's entire dark matter halo as one rigid component (see Section \ref{subsec:orbit_methods}), and therefore, we cannot separate the displacement and reflex motion of M31's inner and outer halo. If we assume that the memory of M31's COM motion is retained by halo stars that have not orbited around M31's COM for more than one dynamical time, we can translate the evolutionary history of M31's COM position and velocity from the orbits presented in Section \ref{sec:orbit_results} to observables of the expected travel velocity (i.e., velocity offset between M31's inner and outer halo).

Here, we quantify M31's expected travel velocity as a function of distance from M31's COM with the aim of finding a specific distance range within M31's outer halo that satisfies two criteria. The first is that the volume of M31's halo corresponding to this distance is densely populated enough to statistically separate M31 stars from MW foreground stars \citep[e.g., see][]{gilbert12, gilbert18}. The second is to identify a distance range where halo stars will have retained the memory of M31's reflex motion in response to M33 (i.e., the signal has not yet washed out). Relatedly, it is important that the magnitude of M31's expected travel velocity at that distance is significantly larger than the typical precision of observable stellar motions, which we return to at the end of this subsection.

Figure~\ref{fig:m31COM} shows that M31's COM position and velocity shift were at maximum at 6~Gyr ago for all M33 mass models. Objects orbiting in our assumed M31 potential at a distance of 250~kpc have a dynamical time of 6~Gyr.
However, the density of stars in M31's halo at 250~kpc is too low compared to the MW foreground ($\mu_{\rm M31, halo} \sim 33$ mag arcsec$^{-2}$; 
\citealt{courteau11,gilbert12}) 
to have an appreciable signature of M33's dynamical impact. Moreover, red giant branch stars in M31's halo have yet to be spectroscopically confirmed beyond a radius of $\sim$175~kpc \citep{gilbert12}. 
To maximize prospects of measuring M31's halo response to M33 while satisfying the criteria above, we focus on stars between 100--150~kpc where M31's halo is more densely populated with red giant branch stars ($\mu_{\rm M31, halo} \sim 31.5$ mag arcsec$^{-2}$ 
at 125~kpc, or a factor of $\sim$4.5 times higher M31 halo star density assuming a constant MW foreground). Stars at these distances correspond to dynamical times of $\approx$1-3~Gyr ago. From Figure~\ref{fig:m31COM}, the corresponding M31 COM position and velocity at 3~Gyr ago is shifted $\lesssim$40~kpc and $\lesssim$25~km s$^{-1}$ from the origin, respectively, for the highest mass M33.

To translate M31's COM velocity shift to a reflex motion, the expected shift between the inner and outer M31 halo, we extract the M31 6D phase space vector corresponding to 3 Gyr ago from the orbit data calculated in Section \ref{sec:orbit_results} (assuming the highest mass M33). We then multiply by the inverse rotation matrix to remove the geometric correction that accounts for the orientation of the disk and translate it back to the Galactocentric frame from the M31-centric coordinate system. The reflex motion is converted to an observable, the predicted travel velocity split into its radial and tangential components in the heliocentric frame for convenience. 

In Table \ref{tab:prediction}, we report the predicted M31 travel velocity assuming M31's disk (plus inner halo) is moving at the line-of-sight motion of M31 ($-$301 km s$^{-1}$) in the radial direction. Velocity offsets can be positive or negative in the radial direction corresponding to whether stars on the nearside or far side of M31 are observed. In reality, it is not observationally feasible to determine the distance to M31 halo stars relative to its disk. For the travel velocity in the tangential direction, we assume M31's disk is moving at $\rm V_{tan}=145 \, km\, s^{-1}$, the tangential velocity zero-point of the HST+sats data set in heliocentric coordinates \citep{vdm12ii}.
 
From the MW's vantage point, the travel velocity is maximized along the radial direction for M33's orbit in a high mass M31 halo. For the low mass M31, the magnitude of travel velocity is almost equivalent along the radial and tangential directions. Precise radial velocities for a statistically significant sample of M31 halo stars at the listed distances are, therefore, the recommended pathway for measuring M31's travel velocity, especially due to the significant observational investment needed to measure tangential velocities via proper motion measurements. 

For a more robust prediction of M33's dynamical imprint on M31's halo, idealized N-body simulations are necessary to capture M31's COM displacement and reflex motion, not just the evolution of M31's COM position and velocity, as a function of radius from the center of the galaxy. Such models would also provide important information about the volume of M31's halo where this signal is maximized (i.e., on the near or far side of M31) and, subsequently, whether a positive or negative shift relative to M31's bulk radial motion is expected.

\subsubsection{Prospects for Observing M31's Travel Velocity}
Line-of-sight velocities in M31's stellar halo\footnote{Here, we specifically refer to the component of M31's stellar halo that is phase-mixed (i.e., free of tidal substructure within detection limits) and far from M31's disk.} have been measured to typical precisions of $\sim$4 km s$^{-1}$ from ground-based spectroscopy of predominantly individual red giant branch stars (e.g., \citealt{raja05,chapman06,gilbert06,kalirai2006,koch08,escala20b,wojno23}). In particular, the Spectroscopic and Photometric Landscape of Andromeda's Stellar Halo (SPLASH) survey has provided the only measurements of radial velocities for M31 halo stars at sufficiently large projected distances (e.g., \citealt{gilbert12,gilbert18}) from M31's center to in principle detect the predicted travel velocity induced by the M31-M33 interaction (Table~\ref{tab:prediction}). Other existing spectroscopic datasets, such as the DESI survey of M31's stellar halo \citep{dey23}, are too centrally concentrated for a plausible detection of the travel velocity, with a typical maximum radial extent of approximately 50 projected~kpc from M31's center. 

However, placing observational constraints on the travel velocity from currently available SPLASH data faces limitations from spatial coverage. 
The current SPLASH footprint in M31's outer halo, which is composed of $\sim$10 sparsely located Keck/DEIMOS fields in the relevant projected radial range (100--150~kpc), could be expanded by a targeted spectroscopic campaign guided by future idealized N-body simulations of the interaction between M33 and M31. Such simulations could capture M31's COM displacement and reflex motion as a function of radius from the center of the galaxy while simultaneously providing important information about the hemisphere of M31's halo where the reflex motion is maximized.

Regardless of spatial coverage, any spectroscopic survey of M31's outer halo must contend with contamination from stars in the foreground of the MW, which overwhelmingly dominate M31 giants in star counts \citep{gilbert12,gilbert18}. The properties of MW and M31 stars, including radial velocities, exhibit significant overlap in their distributions (e.g., \citealt{gilbert06, escala20b}), making it difficult to disentangle the two populations when the distances to such inherently faint stars are unknown. A sophisticated kinematical separation of M31 giant stars from MW foreground stars (e.g., \citealt{gilbert18}) would be required to place robust observational limits on the travel velocity. 
Moreover, the intrinsically low number statistics for stars in M31's outer halo presents a challenge in terms of distinguishing between the phase-mixed component of M31's halo, which is the relevant stellar structure for measuring the travel velocity, and unknown tidal substructures in M31.

Future dedicated spectroscopic surveys targeting red giant branch stars with DESI could improve statistics toward constraining the prevalence of substructure at 100--150 projected~kpc in M31's outer halo and isolating the travel velocity signal. Although a wide-field contiguous spectroscopic survey of M31's halo is planned with the Prime Focus Spectrograph (PFS) on Subaru, it will only extend to $\sim$100 projected~kpc in the direction of M31's Northwest Stream \citep{ogami24}, limiting its ability to constrain the impact of M33 on M31's halo. Finally, future dedicated spectroscopic survey facilities such as the proposed Maunakea Spectroscopic Explorer (MSE) could provide a magnitude-limited spectroscopic census of M31's halo out to one-half its virial radius \citep{gilbert19}, enabling studies of M31's global dynamical history.

\section{Summary and Conclusions}
\label{sec:conclusions}

We have constrained the orbital history of M33 about M31 using recent measurements of their combined 6D phase space information. Orbits have been computed in two M31 potentials, one with a low mass of $\rm M_{vir}= 1.5 \times 10^{12} \, M_{\odot}$ and one with a high mass of $\rm M_{vir}= 3 \times 10^{12} \, M_{\odot}$. We first demonstrated the effects of allowing M31's COM to move in response to M33's gravitational influence compared to a fixed M31 COM. Using the resulting suite of orbital histories, we have determined the effect of M33's passage on the center of mass position and velocity of M31 over the last 6 Gyr. Results are reported assuming two tangential velocities for M31 from the recent literature: one from HST, denoted as HST+sats, and one from \emph{Gaia} denoted as \emph{Gaia} eDR3. This is the first quantification of a massive satellite's dynamical influence on its host galaxy's center of mass outside of the MW. We summarize the main conclusions of this paper below.

\begin{enumerate}\setlength{\leftmargin}{0pt}
    \item As our first exercise, we investigate the consequences of ignoring M33's dynamical impact on M31's COM motion. When M31's COM is artificially fixed at the origin, all M33 orbits exhibit an increased orbital period and typically orbit at larger distances (see Fig.~\ref{fig:fixedvsmovingM31}) than when M31's COM can respond to M33's gravitational influence. Allowing M31's COM to move does not yield an overall different orbital solution than the fixed scenario. However, it does shift the barycenter of the M31-M33 system by up to a few tens of kiloparsecs away from M31's center.

    \item Adopting the HST+sats M31 tangential velocity, M33's orbital history in the low mass M31 potential is preferentially a first infall orbit (statistical likelihood $>$90\%, see Tables \ref{tab:orbit_params1} and \ref{tab:orbit_params2}). For the high mass M31, a pericentric passage is common for $\sim$65\% of M33 orbits, but pericentric distances are typically too large ($>$100~kpc) to cause morphological tidal features. With the \emph{Gaia} eDR3 M31 tangential velocity, a first infall scenario is also the most common orbital history for M33 in both the low (100\% of orbits) and high mass ($\sim$90\% of orbits) M31 potential. See Fig.~\ref{fig:m33orbits}.
    
    \item Using the M33-M31 orbits computed with a moving M31 COM, we quantified the evolution of M31's COM position (\rcom{}) and velocity (\vcom{}) over the last 6 Gyr (see Fig.~\ref{fig:m31COM}). For the HST+sats M31 tangential velocity, \rcom{}=30-150~kpc with minimal differences owing to the choice of M31 mass. However, \vcom{} is more sensitive to whether M33 has completed a pericentric passage, as is the case for the high mass M31 potential. Here, \vcom{}=6-42 km s$^{-1}$ with a peak at pericenter ($\sim$4.5 Gyr ago). For the low mass M31 potential, \vcom{}=6-35 km s$^{-1}$ with a peak at 6 Gyr ago. For the \emph{Gaia} eDR3 M31 tangential velocity, the range of \rcom{} and \vcom{} are relatively the same regardless of M31's total mass such that \rcom{}=18-108~kpc and \vcom{}=4-26 km s$^{-1}$, and both peak at 6 Gyr ago. 
    
    \item M31's \rcom{} and \vcom{} are typically largest along the Z-axis of the adopted M31-centric coordinate system (see Fig.~\ref{fig:m31COM_components}). In other words, \rcom{} and \vcom{} are largest in the direction orthogonal to M31's disk plane, showing promise for measuring an observable signature of M33's impact.
    
    \item For the ten low mass M31 dwarfs where 6D phase space information is available, we examined how 2-body orbits (excluding M33's gravitational influence) are affected by allowing M31's COM to move versus keeping it fixed. For most dwarfs, there are no noticeable changes in the orbits, with the exception of NGC 147 and NGC 185, two of the brighter dwarfs in the sample. For these satellites, allowing M31's COM to move causes the orbital periods of these satellites to decrease by $\sim$5\%. 

    \item When we analyze the combined impact of M33's gravitational influence on these ten dwarfs (i.e., 3-body orbits) in both a moving M31 COM and fixed M31 COM setup, satellite orbits are more significantly perturbed as M33 induces a larger M31 COM displacement from M31, which in turn alters the orbits of its other satellites (see Fig. \ref{fig:dwarf_orbits}). The most significantly affected dwarfs are NGC 147, IC 10, and And~VII, followed by And~XVI, but all ten satellites are impacted by the combined influence of M33 and a moving M31 COM. Notably,  perturbations to the dwarfs' orbits manifest in different ways, including changing the length of the orbital period, the distance at pericenter and/or apocenter, and timing of pericenter and/or apocenter. This further demonstrates M33's gravitational influence is indeed significant and must be taken into account when reconstructing the accretion and evolutionary history of M31 substructures.
    
    \item M31's satellites are distributed in a highly asymmetric configuration, raising the question of what caused this lopsided feature. We compute the orbits of ten M31 satellites backward in a joint M31+M33 potential and then integrate forward, removing the gravitational influence of M33. By comparing the resulting satellite spatial distribution to a scenario where M33's influence is not removed, we find there is a shift in the spatial distribution of satellites but that M33 is unlikely to be the sole driver of the asymmetric satellite configuration (see Fig.~\ref{fig:spatial_distr}). This exercise should be revisited when 6D phase space information is available for the remaining $\sim$20 M31 satellites to determine the extent of M33's influence on the phase space distribution of the M31 satellite system. The primary source of M31’s lopsided satellite distribution, therefore, remains unknown.

    \item Using the M33 orbit derived with the HST+sats M31 \vtan, we translate M31's \vcom{} to an observable signature of M33's impact on M31's COM, namely the expected travel velocity, assuming that M31 halo stars retain the memory of its COM evolution. We focus on M31 halo stars at 100-150~kpc, where dynamical timescales are sufficiently long for the travel velocity to persist to today and where the stellar density is high enough to distinguish M31 stars from MW foreground. From an observer at the Sun, the radial travel velocity of M31 halo stars is between 8-16 km s$^{-1}$ depending on the halo distance and M31 mass (see Table \ref{tab:prediction}). The radial component is predicted to be approximately equal to or significantly larger than the corresponding tangential travel velocity (depending on M31's mass), showing promise for measuring the travel velocity with future spectroscopic surveys of M31's halo. 
    
\end{enumerate}

We note that these results are only a first-order demonstration of the dynamical disequilibrium caused by the most massive satellite orbiting M31 and that detailed N-body simulations are needed to fully quantify the effect of M33's passage on M31 and, subsequently, other halo substructures. Furthermore, the impact of M32, whose evolutionary history remains unknown, may also introduce further disequilibrium, especially if it recently collided with M31's disk \citep[e.g.,][]{dierickx17} or if it is the remnant of a recent major merger \citep{dsouza18}. If the Giant Southern Stream was a separate accretion event from the major merger that left behind M32 as observed today, this event is likely to impact M31's COM similarly.

From the MW system and analogs of the MW in cosmological zoom-in simulations, we have learned that the magnitude of reflex motion and displacement are strongest at the first pericenter \citep{gc24}, so even if there is a non-negligible influence from M32 and the progenitor of the Giant Southern Stream, the impact is likely to be less than the impact of M33 assuming they have completed more than one orbit about M31. With upcoming proper motions for M32 (HST GO 15658, PI: S.T. Sohn), we will test this hypothesis explicitly.

\section{Acknowledgements}
EP is financially supported by NASA through the Hubble Fellowship grant \# HST-HF2-51540.001-A awarded by STScI. STScI is operated by the Association of Universities for Research in Astronomy, Incorporated, under NASA contract NAS5-26555. EP thanks Hayden Foote for providing the \texttt{CONTRA} models used to build the M31 gravitational potentials. The authors are grateful to Gurtina Besla, Dan Weisz, and Alessandro Savino for useful discussions that helped improve the quality of this paper. IE acknowledges support from programs HST GO-15891, GO16235, and GO-16786, provided by NASA through a grant from STScI, which is operated by the Association of Universities for Research in Astronomy, Inc., under NASA contract NAS 5-26555. The authors thank the anonymous referee for suggestions that improved the quality of this paper.

\software{Numpy \citep{numpy},
  SciPy \citep{SciPy-NMeth},
  Matplotlib \citep{matplotlib},
  IPython \citep{ipython},
  Jupyter \citep{jupyter}, 
  Astropy \citep{astropy:2013, astropy:2018, astropy:2022}, 
  Gala \citep{gala}}

\appendix
\section{M33's Orbit in an Updated Low Mass M31 Potential}
\label{sec:appendixA}

In this work, we derive a new low mass M31 potential compared to the one adopted in \citetalias{patel17a}. Both potentials have the same virial mass, but the bulge and disk parameters have been modified as listed in Table \ref{table:m31lowmassparams}.

\begin{table}[t]
\centering
\caption{Parameters used for the analytic representation of M31's low mass potential in \citetalias{patel17a} compared to this paper.}
\label{table:m31lowmassparams}
\begin{tabular}{lcc}\hline\hline 
 & \citetalias{patel17a} & This Work \\ \hline
$\rm M_{vir}$ [10$^{10}$ $\rm M_{\odot}$] & 150 & 150 \\ 
c$_{\rm vir}$ & 9.56 & 9.56 \\ 
$\rm R_{vir}$ [kpc] & 299 & 299 \\
M$_{\rm d}$ [10$^{10}$ $\rm M_{\odot}$] & 8.5 & 7.3 \\ 
R$_{\rm d}$ [kpc] & 5.0 & 5.0 \\
z$_{\rm d}$ [kpc]  & 1.0 & 0.6 \\ 
M$_{\rm b}$ [10$^{10}$ $\rm M_{\odot}$] & 1.9 & 3.1  \\ 
R$_{\rm b}$ [kpc] & 1.0 & 1.0 \\ \hline
\end{tabular}
\end{table}

\begin{figure}[t]
    \centering
    \includegraphics[width=0.45\textwidth]{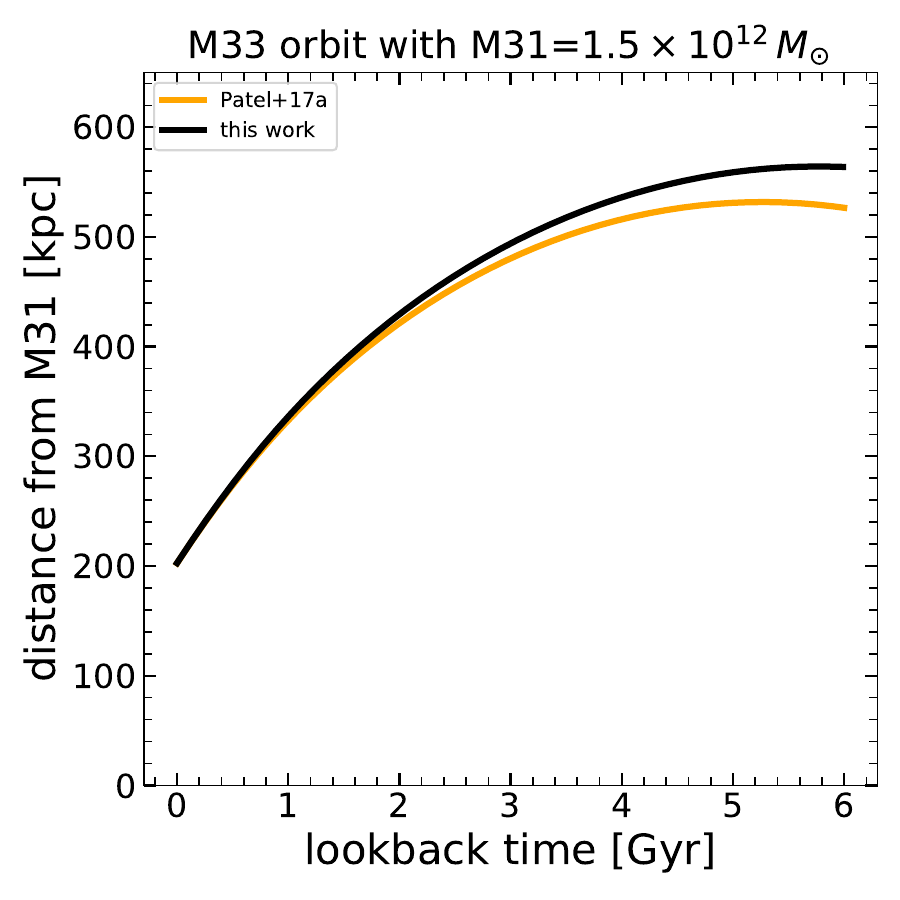}
    \caption{The orbital history of M33 ($M=10^{11}\, M_{\odot}$) in the low mass M31 potential ($\rm M_{vir}= 1.5 \times 10^{12} \, M_{\odot}$) used in this work compared to the low mass M31 potential ($\rm M_{vir}= 1.5 \times 10^{12} \, M_{\odot}$) used in \citetalias{patel17a}. Differences between the two potentials include varying choices of disk and bulge parameters see Table \ref{table:m31lowmassparams})}. We adopt 6D phase space coordinates identical to those used in \citetalias{patel17a} to illustrate that the resulting M33 orbit is very similar in both potentials.
    \label{fig:p17a_comp}
\end{figure}

Figure \ref{fig:p17a_comp} illustrates the resulting direct orbital history for M33 with the new low mass M31 potential used in this work (black) compared to the results of \citetalias{patel17a} (yellow). For consistency, we adopted the M31-M33 coordinates used in \citetalias{patel17a} (see Table 1). The orbit of M33 is very similar using both low mass M31 models, which only differ in the choice of disk and bulge parameters. In both cases, M33 is on first infall, reaching a distance of 500-550~kpc at 6 Gyr ago.

\section{Uncertainties in M31's Displacement and Reflex Motion}
\label{sec:appendixB}

Using the 1,000 orbital histories computed for each M31-M33 mass combination, we quantify the uncertainty on the magnitude of M31's \rcom{} and \vcom{} as a function of time following the same procedure as in Section \ref{sec:orbit_results} for M33. Tables \ref{tab:m31_disp_errors_hst} and \ref{tab:m31_disp_errors_gaia} illustrate the distribution of \rcom{} (second column) and \vcom{} (third column) for the low mass and high M31 potential combined with the highest M33 mass model. Each row corresponds to a specific lookback time. Square brackets list the [15.9, 50, 84.1] percentiles of these distributions.  

The magnitude of \rcom{} is much larger than the corresponding distance uncertainties ($\sim10-15$~kpc per position component) for galaxies in the M31 system \citep[see][]{savino22}. This implies the COM position of M31 satellite galaxies is precise enough to quantify the offset between these galaxies and the COM of M31. The magnitude of \vcom{}, on the other hand, is approximately proportional to or even smaller than the level of uncertainty in the 3D velocities of M31 satellite galaxies on average (typically 20-30 km s$^{-1}$ per velocity component for the most precise measurements. 

For the low mass  M31 potential, the HST+sats results for both \rcom{} and \vcom{} exhibit a much broader range at fixed lookback time, corresponding to the broad diversity of orbital solutions allowed by these 6D phase space uncertainties. M33 can have either a first infall orbit or a recent pericentric passage, whereas the low mass M31 potential with the $Gaia$ eDR3 tangential velocity unanimously results in a first infall orbit. As a result, M31's COM response varies accordingly. The same trend holds for the reflex motion. 

For the high mass M31 potential, the range of \rcom{} and \vcom{} is generally broader than the low mass M31 results, regardless of the assumed M31 tangential velocity. With both data sets, a recent pericentric passage is possible in the high mass M31 potential, though with different statistical significance. As shown in Tables \ref{tab:orbit_params1} and \ref{tab:orbit_params2}, $\sim65$\% of orbits with the HST+sats data have a recent pericenter, whereas only $\sim$11-13\% of M33 orbits have a recent passage. 

Using simulations of LG analogs, \citet{salomon23} showed that velocity offsets (their equivalent of reflex motion) of $\leq$ 10 km s$^{-1}$ are possible in the absence of massive accretion events, but all reflex motion uncertainties reported in Tables \ref{tab:m31_disp_errors_hst} and \ref{tab:m31_disp_errors_gaia} are generally $\geq$ 10 km s$^{-1}$, in good agreement with results from cosmological environments with massive satellite galaxies.

\begin{deluxetable*}{ccc}
\tablecaption{M31 COM Position and Velocity Evolution with HST+sats $v_{\rm tan}$ \label{tab:m31_disp_errors_hst}}
\tablewidth{0pt}
\tablehead{
\colhead{time}  &  \colhead{\rcom{}} & \colhead{\vcom{}} \\
\multicolumn1c{(Gyr)} & \colhead{(kpc)} &  \colhead{(km s$^{-1}$)} }
\startdata
\multicolumn{3}{c}{low mass M31}\\  \hline
1.0 Gyr ago &  [6.80, 7.61, 8.58] & [11.63, 13.30, 15.65] \\
2.0 Gyr ago & [21.10, 24.46, 29.30] & [16.44, 19.94, 25.32] \\
3.0 Gyr ago & [38.99, 46.42, 58.13] & [19.11, 24.19, 32.75] \\
4.0 Gyr ago & [58.94, 72.08, 93.17] & [20.78, 27.45, 39.56]\\
5.0 Gyr ago &  [80.27, 100.59, 134.30] & [22.00, 29.98, 46.07]\\
6.0 Gyr ago &  [102.80,  131.12, 180.32] & [22.94, 32.23, 48.67] \\ \hline
\multicolumn{3}{c}{high mass M31}\\  \hline 
1.0 Gyr ago & [6.97, 7.82, 8.90] & [12.11, 14.00, 16.82] \\
2.0 Gyr ago & [22.31, 26.21, 32.73] & [18.08, 22.68, 31.47] \\
3.0 Gyr ago &  [42.46, 52.21, 69.17] & [22.09, 30.06, 45.08]\\
4.0 Gyr ago & [66.03, 84.36, 105.7] & [25.51, 36.93, 42.76] \\
5.0 Gyr ago & [92.53, 120.42, 133.48] & [28.78, 35.78, 39.87] \\
6.0 Gyr ago & [122.81, 147.79, 156.83] & [28.60, 31.61, 38.72]\\ \hline
\enddata
\tablecomments{The uncertainty on the magnitude of M31's COM position and velocity as a function of time for the highest mass M33 model ($2.5 \times 10^{11}\,M_{\odot}$). At each time, we report the [15.9, 50, 84.1] percentiles on \rcom{} and \vcom{}.}
\end{deluxetable*}

\begin{deluxetable*}{ccc}[h]
\tablecaption{M31 COM Position and Velocity Evolution with $Gaia$ eDR3 $v_{\rm tan}$ \label{tab:m31_disp_errors_gaia}}
\tablewidth{0pt}
\tablehead{
\colhead{time}  &  \colhead{\rcom{}} & \colhead{\vcom{}} \\
\multicolumn1c{(Gyr)} & \colhead{(kpc)} &  \colhead{(km s$^{-1}$)} }
\startdata
\multicolumn{3}{c}{Low Mass M31}\\  \hline 
1.0 Gyr ago & [5.90,  6.53, 7.25] & [ 9.57, 10.78, 12.21] \\
2.0 Gyr ago & [17.22, 19.49, 22.24] & [12.63, 14.66, 17.34] \\
3.0 Gyr ago & [30.68, 35.20,  41.19] & [14.08, 16.78, 20.3 ] \\
4.0 Gyr ago & [45.20,  52.59, 62.44] & [15.01, 18.1,  22.39] \\
5.0 Gyr ago & [60.43, 71.10,  85.41] & [15.64, 19.01, 23.96] \\
6.0 Gyr ago & [76.20,  90.32, 109.9] & [16.06, 19.71, 25.2 ] \\ \hline
\multicolumn{3}{c}{High Mass M31}\\  \hline 
1.0 Gyr ago & [6.00,   6.67, 7.43] & [ 9.87, 11.18, 12.76] \\
2.0 Gyr ago & [17.91, 20.40,  23.58] & [13.44, 15.91, 19.17] \\
3.0 Gyr ago & [32.43, 37.75, 45.08] & [15.40,  18.97, 24.06] \\
4.0 Gyr ago & [48.34, 57.79, 71.17] & [16.78, 21.21, 28.81] \\
5.0 Gyr ago & [65.49,  79.67, 102.39] & [17.70,  23.07, 34.56]\\
6.0 Gyr ago & [83.62, 103.65, 138.05] & [18.47, 24.75, 35.92] \\ \hline
\enddata
\tablecomments{Same as Table \ref{tab:m31_disp_errors_hst} but for the $Gaia$ eDR3 $v_{\rm tan}$ zero point.}
\end{deluxetable*}

\bibliography{postdoc_refs}{}
\bibliographystyle{aasjournal}
\end{document}